\documentclass[fleqn,usenatbib]{mnras}

\usepackage[T1]{fontenc}

\DeclareRobustCommand{\VAN}[3]{#2}
\let\VANthebibliography\thebibliography
\def\thebibliography{\DeclareRobustCommand{\VAN}[3]{##3}\VANthebibliography}

\usepackage{graphicx}	% Including figure files
\usepackage{amsmath}	% Advanced maths commands
\usepackage{amssymb}	% Extra maths symbols
\usepackage{newtxtext,newtxmath}
\usepackage[usenames,dvipsnames]{xcolor}
\usepackage{balance}
\usepackage{orcidlink}

\hypersetup{
    colorlinks = true,
    citecolor = {MidnightBlue},
    linkcolor = {BrickRed},
    urlcolor = {BrickRed}
}

\newcommand*\dd{\mathop{}\!\mathrm{d}}

\newcommand*\Jones{\boldsymbol{\mathcal{J}}}
\newcommand*\coh{\boldsymbol{\mathcal{C}}}
\newcommand*\vis{\boldsymbol{\mathcal{V}}}

\title[High-dimensional inference of radio interferometer beam patterns]{High-dimensional inference of radio interferometer beam patterns \\
I: Parametric model of the HERA beams}

\author[M. J. Wilensky et al.]{Michael J. Wilensky$^{1}$\thanks{E-mail: \url{michael.wilensky@manchester.ac.uk}}\,\orcidlink{0000-0001-7716-9312},
Jacob Burba$^{1}$\,\orcidlink{0000-0002-8465-9341},
Philip Bull$^{1,2}$\,\orcidlink{0000-0001-5668-3101},
Hugh Garsden$^{1}$\,\orcidlink{0009-0001-3949-9342},
Katrine A. Glasscock$^{1}$\,\orcidlink{0000-0001-6894-0902},\newauthor
Nicolas Fagnoni${^3}$\,\orcidlink{0000-0001-5300-3166},
Eloy de Lera Acedo${^3}$\,\orcidlink{0000-0001-8530-6989},
David R. DeBoer$^{4}$\,\orcidlink{0000-0003-3197-2294},
Nima Razavi-Ghods${^3}$\,\orcidlink{0000-0003-2930-5396}
\\
$^{1}$Jodrell Bank Centre for Astrophysics, University of Manchester, Manchester M13 9PL, UK\\
$^{2}$Department of Physics and Astronomy, University of Western Cape, Cape Town 7535, South Africa\\
${^3}$Cavendish Astrophysics, University of Cambridge, Cambridge CB3 0HE, UK\\
${^4}$Department of Astronomy, University of California, Berkeley, CA, 94720-1234, US
}
\pubyear{2024}

\begin{document}
\label{firstpage}
\pagerange{\pageref{firstpage}--\pageref{lastpage}}
\maketitle

\begin{abstract}
Accurate modelling of the primary beam is an important but difficult task in radio astronomy. For high dynamic range problems such as 21cm intensity mapping, small modelling errors in the sidelobes and spectral structure of the beams can translate into significant systematic errors. Realistic beams exhibit complex spatial and spectral structure, presenting a major challenge for beam measurement and calibration methods. In this paper series, we present a Bayesian framework to infer per-element beam patterns from the interferometric visibilities for large arrays with complex beam structure, assuming a particular (but potentially uncertain) sky model and calibration solution. In this first paper, we develop a compact basis for the beam so that the Bayesian computation is tractable with high-dimensional sampling methods. We use the Hydrogen Epoch of Reionization Array (HERA) as an example, verifying that the basis is capable of describing its single-element E-field beam (i.e. without considering array effects like mutual coupling) with a relatively small number of coefficients. We find that 32 coefficients per feed, incident polarization, and frequency, are sufficient to give percent-level and $\sim$10\% errors in the mainlobe and sidelobes respectively for the current HERA Vivaldi feeds, improving to $\sim 0.1\%$ and $\sim 1\%$ for 128 coefficients.
\end{abstract}


%%%%%%%%%%%%%%%%%%%%%%%%%%%%%%%%%%%%%%%%%%%%%%%%%%

%%%%%%%%%%%%%%%%% BODY OF PAPER %%%%%%%%%%%%%%%%%%

\section{Introduction}

Large radio interferometer arrays are increasingly being used for high dynamic range (HDR) applications, where the signal of interest is several orders of magnitude fainter than other confounding signals \citep{Paciga2011, Tingay2013, LOFAR, santos2017meerklass, deBoer2017, CHIME2022}. HDR observations place extremely stringent requirements on the precision and fidelity of array calibration; whereas relatively small (e.g. percent-level) errors in calibration may be tolerable for `traditional' applications such as imaging bright sources, they can cause catastrophic leakage and misidentification of bright signals into parts of the data (e.g. particular regions of Fourier space) where we might otherwise have hoped to extract the faint target signal. HDR applications are challenging for standard calibration methods, and a great deal of effort has been put into methodological development to overcome their limitations and provide calibrations at the required level of accuracy \citep{Dillon2020, AEW2022, Byrne2023, Charles2023, Cox2023}.

A prominent example is the detection of brightness temperature fluctuations of the redshifted 21cm line from neutral hydrogen, commonly known as 21cm intensity mapping (IM). In this application, the 21cm fluctuations are expected to be around the $\sim$~mK level, compared with Galactic and extragalactic synchrotron emission (`foregrounds') ranging from tens of Kelvin in brightness temperature at higher frequencies ($\sim 1$~GHz) to thousands of Kelvin at lower frequencies ($\lesssim 200$~MHz) \citep[see, e.g.][]{mesinger, liu-shaw}. This represents a typical dynamic range of $\sim10^4-10^5:1$. In total intensity (Stokes I), the foreground emission is expected to have a smooth, power-law-like frequency spectrum, as compared with the rapidly fluctuating spectrum of the 21cm fluctuations, which in principle should allow the signals to be separated using standard tools such as Fourier filtering, Principal Component Analysis, and the like \citep{morales-hewitt, Morales2012}. Calibration errors and artifacts greatly complicate this picture however; even relatively small errors can couple bright, foreground-dominated Fourier modes into the 21cm signal-dominated modes, swamping the signal \citep{Orosz_2019, Barry2016, Byrne2019}. We require instrumental calibrations at a level better than the dynamic range, of order $10^{-5}$, to keep the leaked foregrounds below the 21cm signal \citep{Barry2016, Thyagarajan2016}. This is likely to require knowledge of the instrument or a reference sky model at least a couple of orders of magnitude better than what is currently possible \citep{Shaw2015, Ewall-Wice2017}.

While there are many aspects of an interferometer array that need calibration, these can largely be combined into two types of complex gain parameters -- {\it direction-independent} gains, which describe the frequency- and time-dependent degrees of freedom such as the bandpass, and {\it direction-dependent} gains, which incorporate the change in sensitivity with angle due to each receiving element's antenna pattern (beam), also as a function of frequency and time. Direction-independent gain calibration has been studied extensively in the context of 21cm IM experiments {\citep[e.g.][]{Byrne2019, Byrne2021, Byrne2023, Charles2023, Dillon2020, Ewall-Wice2017, AEW2022}}. In this paper we focus on direction-dependent calibration, and in particular inference of the per-antenna complex (E-field) beam.

Beam estimation is difficult because the beam itself has a large dynamic range, a complicated dependence on direction and frequency, and multiplies a sky that is also subject to substantial modelling uncertainties. Even antennas that are designed to have large directivity, such as the parabolic dish reflectors of the Hydrogen Epoch of Reionization Array (HERA), exhibit non-negligible sidelobes far from the centre of the beam pattern. The (peak-normalised) maximum sidelobe level of the HERA Vivaldi feed design is in the region of $-15$~dB to $-25$~dB depending on frequency, for instance \citep{Fagnoni2021b}, with substantial variation with frequency. Modelling these sidelobes accurately is a necessity if one hopes to reach a dynamic range of $10^{-4}$ (i.e. $-40$~dB) or better. The beam pattern is also expected to vary somewhat between different elements of the array, as slight variations in feed position and alignment, reflector geometry etc. will always occur \citep{Orosz_2019, Choudhuri+21, Kim2022, Kim2023}. These variations may be exacerbated by environmental conditions, for example different wind loadings or air temperatures at different positions within the array. Electromagnetic models and lab-based beam measurements (which usually cannot be performed in the far field) are typically unable to account for these complicated variations in a robust manner, and so measuring the beam patterns of the array elements in situ is necessary.

A variety of approaches to in situ measurements have been tried. One is to keep track of the autocorrelation (zero-spacing) signal as the sky rotates through the beam as it points in a fixed direction. Using a sky model, one can try to reconstruct a model for the beam that accounts for the antenna temperature variations as different parts of the sky (e.g. bright sources) drift through different parts of the beam. This is generally difficult, as sky models are neither complete nor calibrated at the required level of accuracy. A similar approach uses the visibilities from pairs of correlated antennas, again with a sky model being used as a reference \citep{2012AJ....143...53P, 2020ApJ...897....5N}. The two antenna beams modulate one another however, and other aspects of the interferometer (such as direction-independent gains) compound the measurements, in addition to the familiar issue of the sky model being incomplete. To try and get around the incompleteness problem, attempts have also been made to use bright artificial sources such as drone-mounted radio transmitters to map the beams \citep[e.g.][]{2014IAWPL..13..169V, 2015ExA....39..405P, echo}, or bright emission from satellites \citep[][\textcolor{MidnightBlue}{Chokshi et al. in prep.}]{Line2018}. Other methods such as holography and a combined photogrammetry plus detailed EM modelling approach have also been attempted \citep{Berger2016, VLA2019}. %\textcolor{red}{Probably want some citations for some of these. Not that the reader will doubt the credibility of the statements, but more so that they can go read about these topics if they want.}

In this paper series, we describe and demonstrate a parametric Bayesian method for inferring the per-element beam from observed visibilities. Bayesian instrumental characterization methods have been demonstrated before in the context of radio astronomy \citep[e.g.][]{Yatawatta2018, Lochner2015, BayesCal1, BayesCal2, anstey2023, cumner2023, Sims2023}. In this paper (Paper I), we focus on the model-building aspect of the inference, and in the second paper (Paper II) we demonstrate the feasibility of the inference on a toy problem. In particular, we build a flexible linear model based on a set of analytic basis functions that describe the simulated E-field beams of the HERA receiving elements with relatively few free parameters. An accurate and efficient basis that respects the symmetries of the problem is essential for avoiding significant model errors that could introduce spurious spectral and spatial structure. Achieving this in a compact way (i.e. with the fewest basis functions/coefficients possible) improves the prospect of directly inferring the structure of the beams from observations -- particularly if this is to be done for each antenna within the array (rather than as an array average), out to the far sidelobes, and/or as a function of frequency. Conversely, a more 'cautious', overly flexible basis would be harder to constrain with typical data, essentially only allowing us to marginalise over a very broad range of possible beam structures. The basis we propose in this paper has many of the properties required for a successful inference scheme in which individual antenna patterns can be measured versus frequency, and is tuneable in terms of its precision, at least with respect to the complex EM simulations that we use as a reference. Note that the methods we use in Paper I to determine this `sparse basis' are not themselves Bayesian; the basis is in support of a Bayesian Gibbs sampling framework expounded in Paper II.

Analytic parametrization of the beam is an active area of research in the HERA collaboration \citep{Choudhuri+21}.\footnote{ \url{http://reionization.org/manual_uploads/HERA101_Analytic_polarized_beam.pdf}}\footnote{\url{https://reionization.org/manual_uploads/HERA114_MEMO_beam_harmonics_Cynthia.pdf}} In general, analytic parametrizations have demonstrated significant promise in solving problems in HDR applications, e.g. more rapid beam evaluation \citep{Asad2021}, better physical beam characterization \citep{Sekhar2022, Nasirudin2022}, and removal of calibration biases that stem from gridding errors \citep{Barry2022}. This last reference also highlights the importance of accurate beam models for the sake of accurate gridding kernels in imaging power spectrum analyses \citep{Morales2019}. 

We introduce a basis set that, to our knowledge, has not been used in the beam modeling literature. This basis solves several problems for us where other similar solutions from the literature did not. 
\begin{enumerate}
    \item It is fully analytic (unlike principle component analysis of holographic measurements for example). This makes it highly portable between different pixelization schemes while avoiding problems associated with discretization issues (e.g. spurious fluctuations in apparent source brightness at the horizon due to the setting of discrete pixel centers, or issues associated with sampling on sub-pixel scales for arrays with long baselines relative to the beam model resolution). 
    \item It describes the complex, realistic HERA beam simulations with relatively few basis functions compared to alternatives.
    \item It behaves well at a coordinate singularity where e.g. polynomial-based bases did not (see \S\ref{sec:math}).
    \item The special functions involved have numerically stable \texttt{scipy} implementations up to high order which allows for thorough exploration of the parameter space.
\end{enumerate}
We implement the sparse basis construction code of Paper I as well as the Gibbs sampling code of Paper II as a part of the \textsc{Hydra} Gibbs sampling framework \citep{Kennedy2023}.\footnote{\url{https://github.com/HydraRadio/Hydra}} The sparse fitting code makes use of the \texttt{UVBeam} class in the \texttt{pyuvdata}\footnote{\url{https://github.com/RadioAstronomySoftwareGroup/pyuvdata}} package \citep{Hazelton2017}. This class provides a general python interface for a variety of beam file formats, meaning the code we provide can be readily applied to other experiments.

% Novel things in this paper
% 1. Newly explored basis set for beam modeling that is useful for az/za polarization convention
% 2. 
%

In what follows, we specialise to interferometric arrays operating in a drift-scan configuration (i.e. always pointing at the zenith), similar to HERA. Using previous EM simulations of the HERA  Phase I and Vivaldi feeds, we show how an efficient basis can be constructed that allows accurate modelling of uncertain E-field beams with a relatively small number of parameters (\S\ref{sec:math}-\ref{sec:basis}). We then explore whether any spurious spectral structure arises as a complication of choosing such a sparse spatial basis (\S\ref{sec:spectral}). Lastly, we conclude (\S\ref{sec:conclusion}).% In the following paper, we demonstrate inference of the beam parameters using simulations with a fixed sky model, but then show how this approach can be incorporated within a Gibbs sampling framework that also allows marginalisation over an uncertain sky model.

%The paper is organised as follows. In Sec.~\ref{sec:math}, we present the general mathematical formalism in which our beam model is embedded. In Sec.~\ref{sec:basis} we examine different approaches to `compressing' the set of basis functions used to model the beam in order to reduce the number of free parameters in the model as far as is practical.  In Sec.~\ref{sec:spectral}, we investigate the possibility of spurious spectral structure introduced by our choice of basis.  Lastly, we conclude in Sec.~\ref{sec:conclusion}.

\section{Mathematical Formalism}
\label{sec:math}

In order to incorporate the beam into our Bayesian inference, we will need to posit a parameterized model for the beam that relates to the data from which the inference is drawn. In the general landscape of Bayesian inference, this is a rather arbitrary decision moderated only by the fact that some parametrizations may describe the data more efficiently than others. Without extreme prior knowledge about the beam, and in particular, uncertainties in the beam due to physical variations from the ideal scenario, we are left to choose some arbitrary basis for the beam with the hopes that our existing prior knowledge can at least constrain which basis functions are the most dominant. As a concrete example, we can take a detailed electromagnetic simulation or highly precise holographic measurement and fit a linear basis to the beam pattern for a high number of modes. Depending on the beam pattern and choice of basis functions, we may find that most of the fit is dominated by a small subset of basis functions. This has been shown to be relatively effective in \citet{deLeraAcedo13, Young2013, bui2017, VLA2019, Asad2021}, though the choice of basis is often idiosyncratic to the telescope. We can then restrict ourselves to these dominant basis functions and consider the coefficients as the free parameters in our model.

\begin{figure*}
    \centering
    \includegraphics[width=\linewidth]{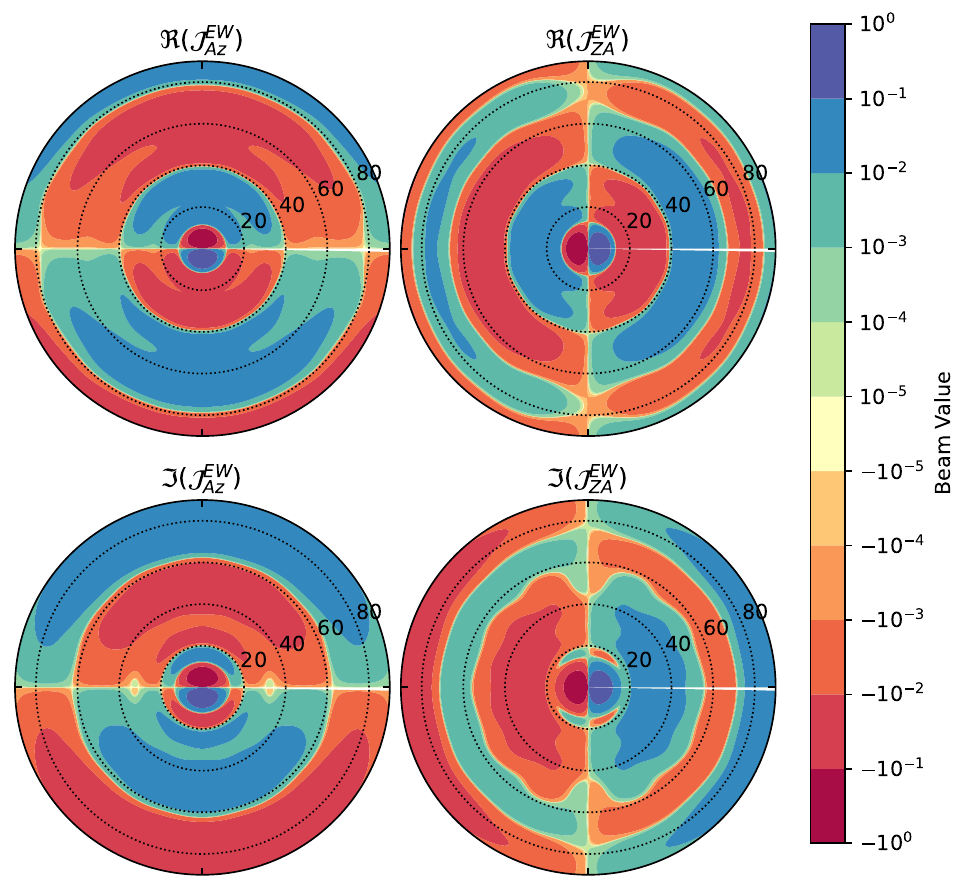}
    \caption{Jones components of the simulated Phase I E-W aligned dipole at 150 MHz, separated into real and imaginary components (rows), and by incident polarization (columns). The symbol $\mathcal{J}^\mathrm{EW}_\mathrm{Az}$ indicates the Jones component for the East-West aligned dipole in response to radiation polarized along the azimuth, while $\mathcal{J}^\mathrm{EW}_\mathrm{ZA}$ indicates the Jones component in response to radiation polarized along the zenith angle. There is an obvious dipolar structure, and higher azimuthal modes are clearly visible at zenith angles of 60 degrees or greater.}
    \label{fig:dipole_beam_150}
\end{figure*}

An appealing alternative, though one that seems to us extremely computationally demanding for a full Bayesian computation, would be to have not only detailed knowledge of the ideal receiving element, but also have a parameterization for physical perturbations to this ideal along with detailed simulations or calculations of the response to them. Then a few parameters describing the perturbations only need to be constrained. See, for example, the characteristic basis function pattern approach in \citet{Maaskant2012, Young2013}. The primary issue is that perturbations tend to belong to a multidimensional continuum and developing simulation outputs for finely graded perturbations to understand the resulting effect on the beam pattern is computationally prohibitive, though not infeasible depending on the study design \citep{Kim2022, Kim2023}. If a suitable emulator were constructed (i.e. if we could apply machine learning to produce fast but necessarily incomplete outputs of the simulator), the computational overhead could be reduced in the long term with an initial investment. However this might violate intellectual property law and is a somewhat ill-defined problem for a general electromagnetic simulator. Due to these seemingly intense challenges, and since our goal is mainly to establish the tractability and usefulness of our Gibbs sampling method, we opt for a more arbitrary linear basis expansion.

\begin{figure*}
    \centering
    \includegraphics[width=\linewidth]{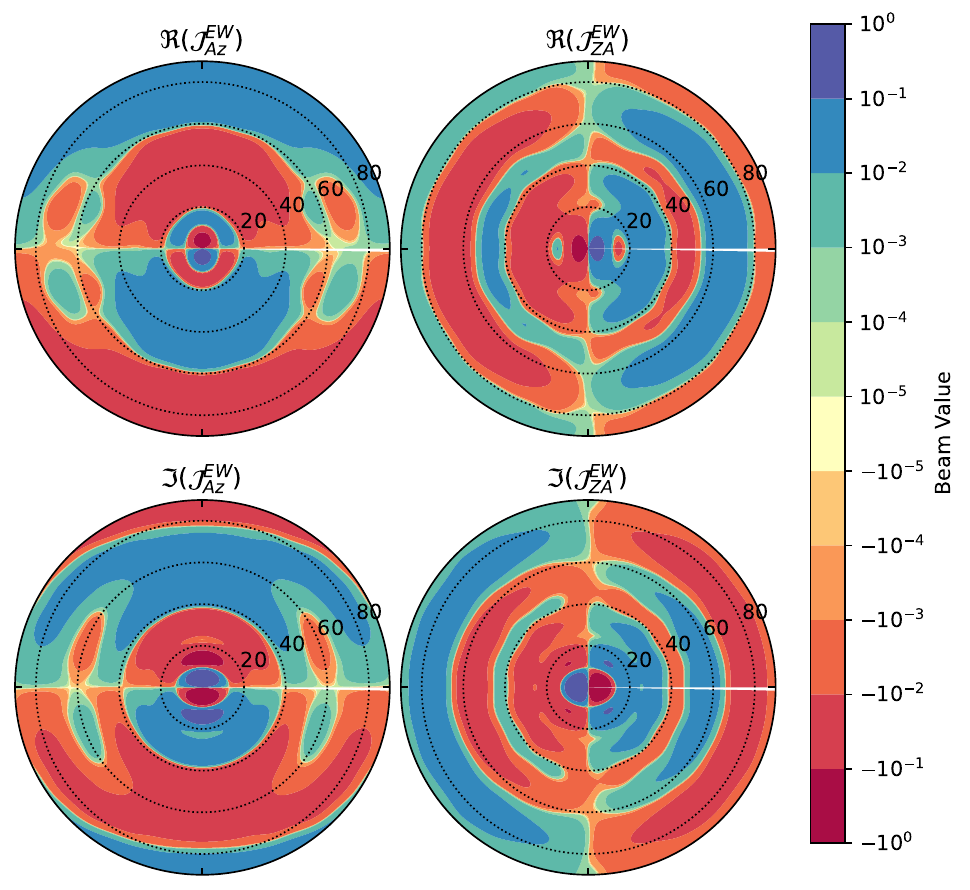}
    \caption{Jones components of simulated HERA vivaldi feed at 150 MHz. Component notation and layout is identical to Figure \ref{fig:dipole_beam_150}. Spatial structure is somewhat more complex than the Phase I dipoles.}
    \label{fig:vivaldi_beam_150}
\end{figure*}

We aim to infer the beam parameters from the visibilities via a forward model. We write the visibility equation in terms of the polarized beam response for antenna $j$ (Jones matrix, ignoring other propagation effects such as ionospheric refraction), $\Jones_j(\hat{s}, \nu)$, sky coherency matrix, $\coh(\hat{s}, \nu)$, and physical antenna separation for antennas $j$ and $k$, $\Delta\vec{x}_{jk}$, as,
\begin{equation}
    \vis_{jk}(\nu, t) = \int \dd\hat{s} \Jones_j\coh\Jones_k^\dag \exp(2\pi i \hat{s}\cdot\Delta\vec{x}_{jk} \nu / c),
\end{equation}
where $c$ is the speed of light, $\nu$ is the frequency of observation, $\hat{s}$ is a unit vector pointing to different positions on the sky, $i^2=-1$, and depending on the coordinate system, some of these quantities are time-dependent. Different choices of coordinate system present different trade-offs. The Bayesian inference for the beam is drastically simpler if we operate in a coordinate system where the beam pattern is fixed. For the sky coordinates, we choose azimuth, $\phi$ and zenith angle, $\theta$. We choose sky polarization directions that align with this coordinate system. This puts all of the time dependence in the sky coherency matrix at the cost of producing a discontinuity in the beam pattern at zenith. This discontinuity arises because in this coordinate system, there is no well-defined direction for the polarization unit vectors at zenith. In math, if we think of the polarization unit vectors, $\hat{\theta}$ and $\hat{\phi}$ as vector fields, then 
\begin{equation}
    \lim_{\theta\rightarrow 0} \hat{\theta}(\theta, \phi_0) \neq \lim_{\theta\rightarrow 0} \hat{\theta}(\theta, \phi_1)
\end{equation}
for $\phi_0 \neq \phi_1$, and similarly for the azimuthal unit vector. See \citet{Byrne2022} for an illustration of this coordinate-based phenomenon, as well as an excellent discussion of state-of-the-art polarized imaging techniques. 

For a pure drift scan observing strategy, this is relatively simple to deal with. In particular we can choose a basis that naturally incorporates this discontinuity. We single out a Jones element and write it as
\begin{equation}
    \mathcal{J}_{jp'}^p(\theta, \phi) = \sum_{n}\sum_{m}a_{jp'nm}^p f_n(\theta)g_m(\phi),
\end{equation}
where $p$ indexes instrumental polarization and $p'$ indexes incident polarization of incoming radiation. If we can find a complete, orthogonal basis such that $f_n(0) \neq 0$ for all $n$, then every single basis function will have this discontinuity at zenith. This requirement makes it such that a best-fit function for some finite number of basis modes does not fill in the discontinuity with 0, as we have seen happen when using Zernike polynomials on projections of the sky to the unit disc.\footnote{Oddly, several instances in the literature claim success with Zernike polynomials \citep{Asad2021, Sekhar2022}. This may just be a result of the polarization basis.} 

\begin{figure*}
    \centering
    \includegraphics[width=\linewidth]{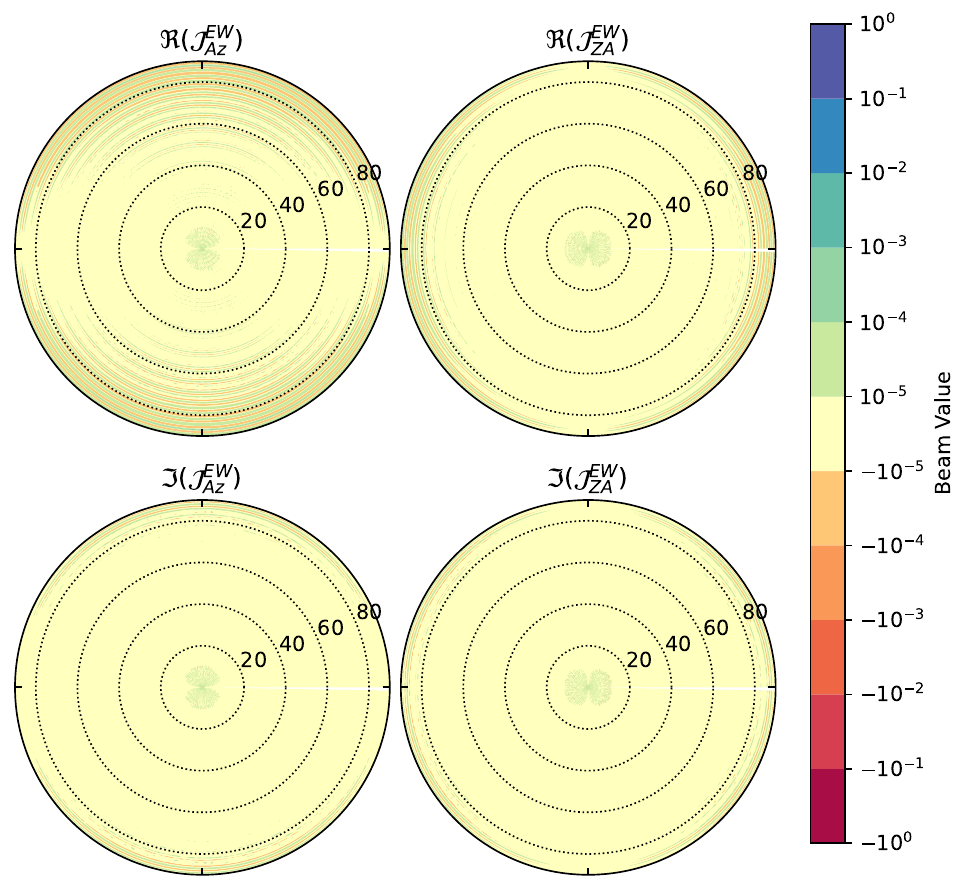}
    \caption{Phase I Dipole residuals for the EW polarization response at 150 MHz using a fit with 80 radial modes and 91 azimuthal modes (7280 total coefficients). Almost all errors are below $10^{-4}$. Errors appear worse in the response to azimuthal polarization.}
    \label{fig:dip_resid}
\end{figure*}

Fortunately there exists such a basis that is fairly natural to use with a dipole radiator. If we project the sky to the unit disc, we can use a Fourier-Bessel series to capture the radiation pattern. From our observations, the azimuthal (Fourier) modes capture the azimuthal structure of the beam patterns in question with relatively few modes. The basis can be made complete and contain the discontinuity in every basis function by using $0$th order Bessel modes and demanding that a linear combination of the basis function and its derivative go to 0 at the horizon \citep{jackson}. Generally, we write
\begin{equation}
    \mathcal{J}_{jp'}^p(\theta, \phi) = \sum_{n}\sum_{m}a_{jp'nm}^p \frac{J_0(u_n\rho(\theta))e^{im\phi}}{q_n},
\end{equation}
where $u_n$ is the $n$th root of the linear combination of Bessel functions specifying the boundary condition. If we just demand that either the basis function or its derivative vanishes, then $u_n$ is the $n$th zero of either the $0$th order Bessel function $J_0(x)$ or its derivative, $J_0'(x) = -J_1(x)$ (depending on choice of boundary condition), and $q_{n}$ is a choice of normalization for the basis so that the modulus square of each basis function integrates to 1 over the disc. Since the beam response at the horizon for azimuthally polarized radiation is nonzero, the choice of Neumann (derivative-vanishing) boundary condition seems more suitable there, however in our numerical experiments we find a better overall fit from the Dirichlet (function-vanishing) boundary condition for a finite sum of basis functions. The normalization constant is given explicitly by
\begin{equation}
   q_{n}= \begin{cases}
        \sqrt{\pi}[J_1(u_n)]\text{, Dirichlet boundary conditions} \\
        \sqrt{\pi}[J_2(u_n)]\text{, Neumann boundary conditions}.
    \end{cases}
\end{equation}

\begin{figure*}
    \centering
    \includegraphics[width=\linewidth]{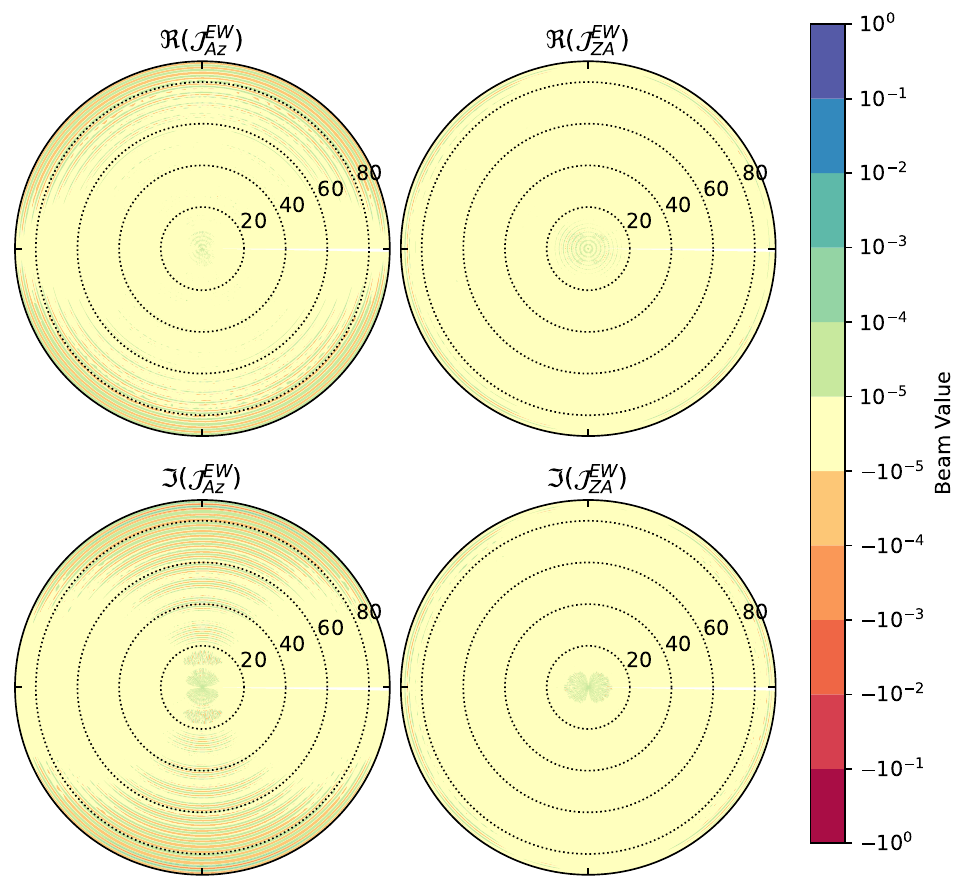}
    \caption{Vivaldi residuals for the EW polarization response at 150 MHz using a fit with 80 radial modes and 91 azimuthal modes (7280 total coefficients). Again, errors are mostly below $10^{-5}$, and are worse in the response to azimuthally polarized radiation.}
    \label{fig:viv_resid}
\end{figure*}

For this approach, we need to pick a projection to the unit disc. While an orthographic projection is a choice with geometrically transparent properties, we instead opt for a projection defined by
\begin{equation}
    \rho = \frac{\sqrt{1 - \cos \theta}}{\alpha},
\end{equation}
where $\alpha$ is close to 1. This projection has the useful property that
\begin{equation}
    \rho \dd\rho \dd\phi = \frac{1}{2\alpha}\sin\theta \dd\theta \dd\phi.
\end{equation}
In other words the projection is area-preserving (up to an overall factor of $2\alpha$). For $\alpha=1$, $\rho=0$ maps to zenith and $\rho = 1$ maps to the horizon. The total volume contributed by a basis function to the square of a Jones element is simply given by
\begin{equation}
    V_{jp'nm}^p = 2\alpha|a_{jp'nm}^p|^2.
    \label{eq:vol}
\end{equation}
As a function of $(n, m)$, we call this the ``Fourier-Bessel Energy Spectrum'' (FBES) for the beam in question.

\begin{figure*}
    \includegraphics[width=\linewidth]{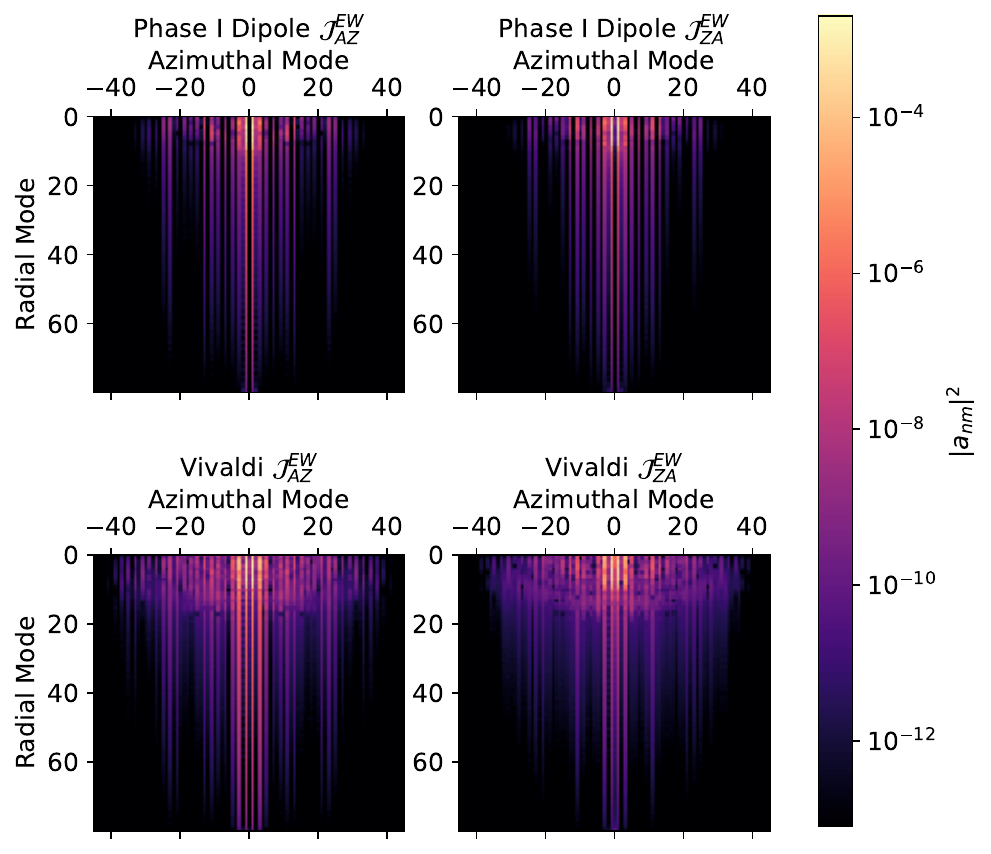}
    \caption{FBES (Equation \ref{eq:vol}) for the Phase I (top row) and Vivaldi (bottom row) feeds and both sky polarizations (left and right) at 150 MHz. The spectra for both feeds contain longer radial tails in the response to azimuth polarization, suggesting more intricate radial structure. This is corroborated by the increased radial structure of the residuals in Figures \ref{fig:dip_resid} and \ref{fig:viv_resid}. The Vivaldi feeds generally more complicated spatial structure compared to the Phase I dipole is exhibited by the larger number of high-energy modes in this plot (noting that both beams have similar volume). }
    \label{fig:bess_es}
\end{figure*}

Using $\alpha=1$ and Dirichlet boundary conditions produces large errors near the horizon since the electromagnetic response is nonzero there. Using Neumann boundary conditions allows us to model the horizon better at the expense of significantly worse fits throughout most of the sky. We find that the horizon problem can be better solved by noting that we can essentially extend the beam with zero-padding to zenith angles beyond the horizon since any source beneath the horizon will contribute 0 flux. We then choose $\alpha$ to be slightly greater than 1, so that the vanishing point of the radial Bessel functions is slightly beneath the horizon. This allows for nonzero values at the horizon and thus greatly reduces errors there. This also improves errors over most of the sky since we are no longer forcing the fit to satisfy a boundary condition that is explicitly disobeyed by the simulation. %The tradeoff is that the basis is now orthonormal on a space corresponding to slightly more than a hemisphere, meaning that \ref{eq:vol} is approximate. 

To investigate the effectiveness of this basis for the HERA use case, we fit the simulated Phase I dipole and Vivaldi beams from \citet{Fagnoni2021a, Fagnoni2021b} in a weighted least-squares sense and observe what varying the number of basis functions does to the performance. Mathematically, we write our Jones element as a vector, $\bold{j}_{j\hat{\alpha}}^p$, where each component is the simulated beam evaluated at a pixel on the sky at 1 degree resolution. We then assume
\begin{equation}
    \bold{j} = \bold{B}\bold{a}
\end{equation}
where $\bold{B}$ is the design matrix encoding our choice of basis, and $\bold{a}$ are the coefficients in that basis. We then calculate a linear least-squares solution for when the number of basis functions is finite. This amounts to minimizing the loss function
\begin{equation}
    L = (\bold{j} - \bold{B}\bold{a})^{\dag}(\bold{j} - \bold{B}\bold{a}),
    \label{eq:lstsq}
\end{equation}
This is equivalent to solving for $\bold{a}$ in the normal equations
\begin{equation}
    \bold{B}^\dag\bold{B}\bold{a} = \bold{B}^\dag\bold{j}.
\end{equation}
In this work, we do this independently for each frequency. In \S\ref{sec:spectral}, we examine the spectral properties of these fits. In Paper II, we use our Bayesian prior to enforce spectral smoothness. In subsequent sections, we analyze the fit performance under different choices of allowed radial and azimuthal modes.

\begin{figure*}
    \includegraphics[width=\linewidth]{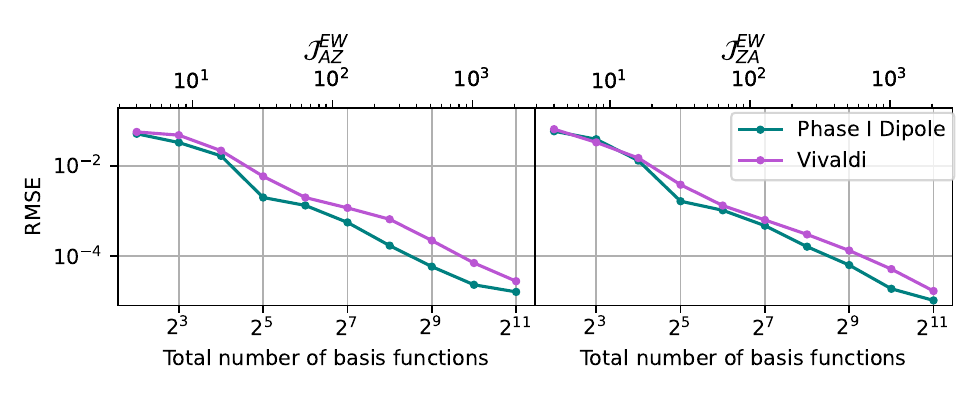}
    \caption{Root-mean-square error (RMSE) of the beam fits at 150 MHz as a function of number of basis functions for each feed and polarization response. Polarization response is separated by panel. The RMSE seems to improve roughly like a power law as we include more basis functions and performance is similar across feed types and polarization. The Vivaldi beam appears slightly less spatially compressible in this basis than the Phase I Dipole, though both seem to obtain good performance with at least $2^7=128$ basis functions.}
    \label{fig:rmse}
\end{figure*}

\section{Compressing the Basis}
\label{sec:basis}

A primary strength of this approach is that we choose a non-pixel basis to express the beam. This means that a set of coefficients can easily map to different pixel schemes without the need for interpolation. However, the coefficients make reference to a particular coordinate system, so this does not eliminate the need for careful coordinate transformations. For the inference, this strength is only manifest if the chosen basis can compactly represent the beam, i.e.\ approximately reproduce the true beam with relatively few coefficients. This is important both from a computational perspective and theoretical inference perspective. We would like to constrain as few parameters as possible given a data set of fixed size, and this problem is magnified by the fact that radio telescope beams have nontrivial frequency structure.  

There are two critical concerns when choosing a basis to represent the beam. First, assuming a priori knowledge of the beam through electromagnetic physics and knowledge of the receiving system, does the basis compactly represent the ideal beam assuming the perfect modeling of the receiving system? Second, assuming a class of perturbations to the receiving system, are ensuing perturbations to the beam pattern also compactly represented by the chosen basis? In this work, we provide a basis that satisfies the first question for the HERA receiving systems (both the Phase I dipole and Vivaldi feeds) using simulations of the ideal receiving system constructed in \citet{Fagnoni2021a, Fagnoni2021b}, which use CST Microwave Studio \citep{Weiland1977, Clemens2001}. We leave a full treatment of the second question for future work.

We show the peak-normalized simulated beams for the HERA Phase I dipole and Vivaldi feeds at 150 MHz in Figures \ref{fig:dipole_beam_150} and \ref{fig:vivaldi_beam_150}. This is the middle of the operating range for both feeds. Generally, the spatial structure of the beam appears more complicated at higher frequencies. We see that the receiving elements have an obvious dipolar pattern that dominates, however more complicated azimuthal structure emerges at larger zenith angles. Since the simulated receiving elements have $90^\circ$ rotational symmetry, the N-S dipole responses are a rotated copy of the E-W polarized responses. In order to increase the visibility of plots for the rest of the paper, we will henceforth only show quantities in terms of the E-W dipole response. In practice, this is not  true, since the full embedded element pattern has its symmetries broken based on where a given receiving element is in the array \citep{Fagnoni2021a, Fagnoni2021b}. We display a proof of concept with the simpler model first, and reserve an analysis of the full embedded element pattern for future work. 

We fit each beam with 80 radial modes and 91 azimuthal modes and show the residuals in Figures \ref{fig:dip_resid} and \ref{fig:viv_resid}. Residuals are generally less than $10^{-5}$, however this is far too many basis functions (7280) to be practical in a Gibbs sampling setting since the coefficients vary as a function of frequency and receiving element. The Phase I dipole residuals appear to be dominated by radial modes all the way to the horizon. The Vivaldi errors are slightly more visually noticeable, but still small. Nevertheless, we are strongly encouraged by the overall size of the errors over most of the sky. 

In order to compress the basis, we need a way of determining which modes are the most important to use. Where appropriate data are available, various statistical methods may be used to perform this compression, such as principle component analysis of holographic measurements \citep{Asad2021}, or by inferring direction-dependent effects in the visibilities using a penalized likelihood \citep{Yatawatta2018}. We are opting to solve this compression problem before setting up the visibility based inference (in Paper II), rather than doing it on line during inference with e.g. a shrinkage estimator \citep{Gelman2013}. To do this, we examine equation \ref{eq:vol} (the FBES) for the very precise fit using thousands of coefficients, shown in Figure \ref{fig:bess_es} for both feed types and sky polarizations at 150 MHz. Almost all of the energy is at low $n$ and $m$ for both feeds, suggesting that we can choose a much smaller basis set with relatively little error. Since all of the plotted Jones elements have similar volume, we can interpret the number of high-energy modes as a proxy for the spatial complexity of the beam i.e. how compressible it is. For both feeds, we see that the response to azimuthally polarized radiation has longer tails to high radial modes compared to the response to zenith angle polarized radiation. Additionally, the Vivaldi feed has more higher energy modes than the Phase I dipole. For example the $|m|=3$ modes are prominent out to fairly high radial power, and there are more significant modes at much higher azimuthal number. 

\begin{figure*}
    \includegraphics[width=\linewidth]{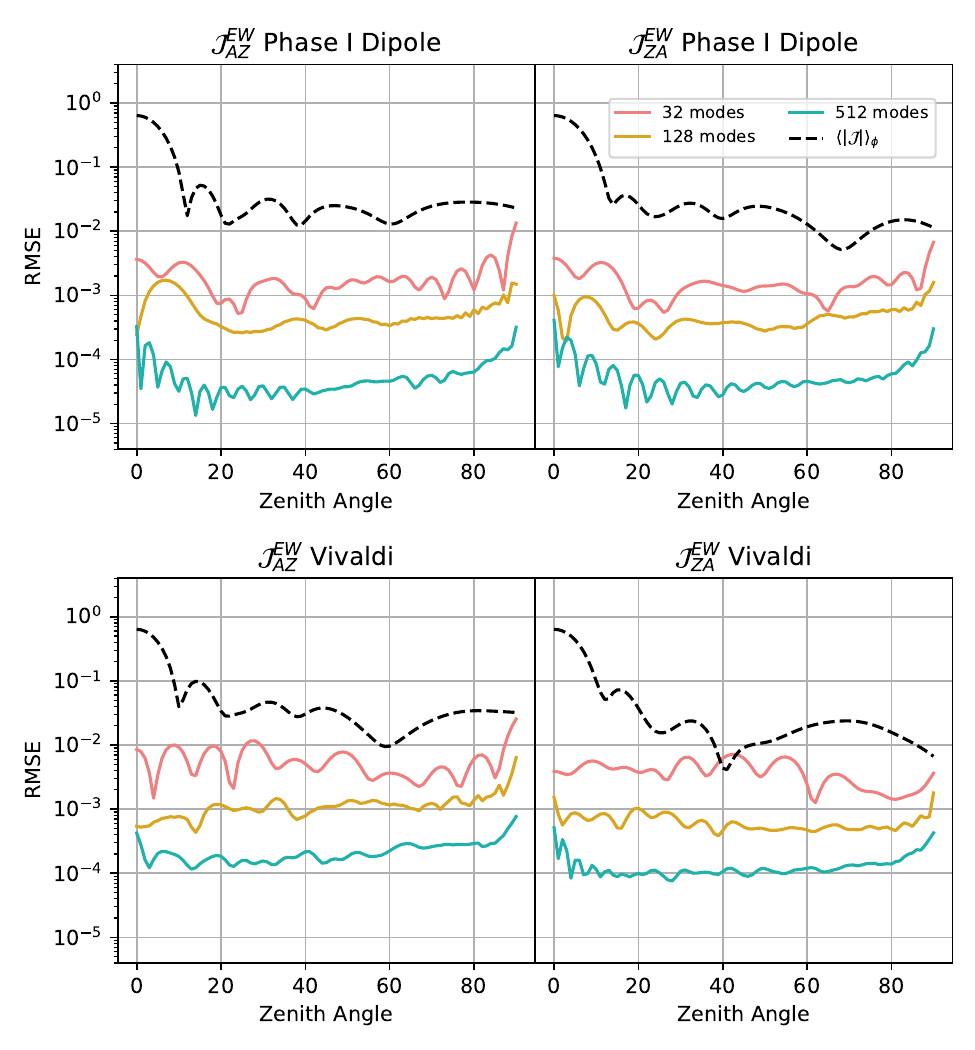}
    \caption{Angular root-mean-square error for each of the feed types and polarizations, along with the average jones element amplitude at each zenith angle. Using just 32 basis functions gives 1\% errors in the main lobe and 10\% errors in the sidelobes for the Phase I feed, and slightly worse for the Vivaldi feed. With 512 modes, we achieve less than 1\% root-mean-square error at almost all zenith angles.}
    \label{fig:ang_rms}
\end{figure*}

\begin{figure*}
    \includegraphics[width=\linewidth]{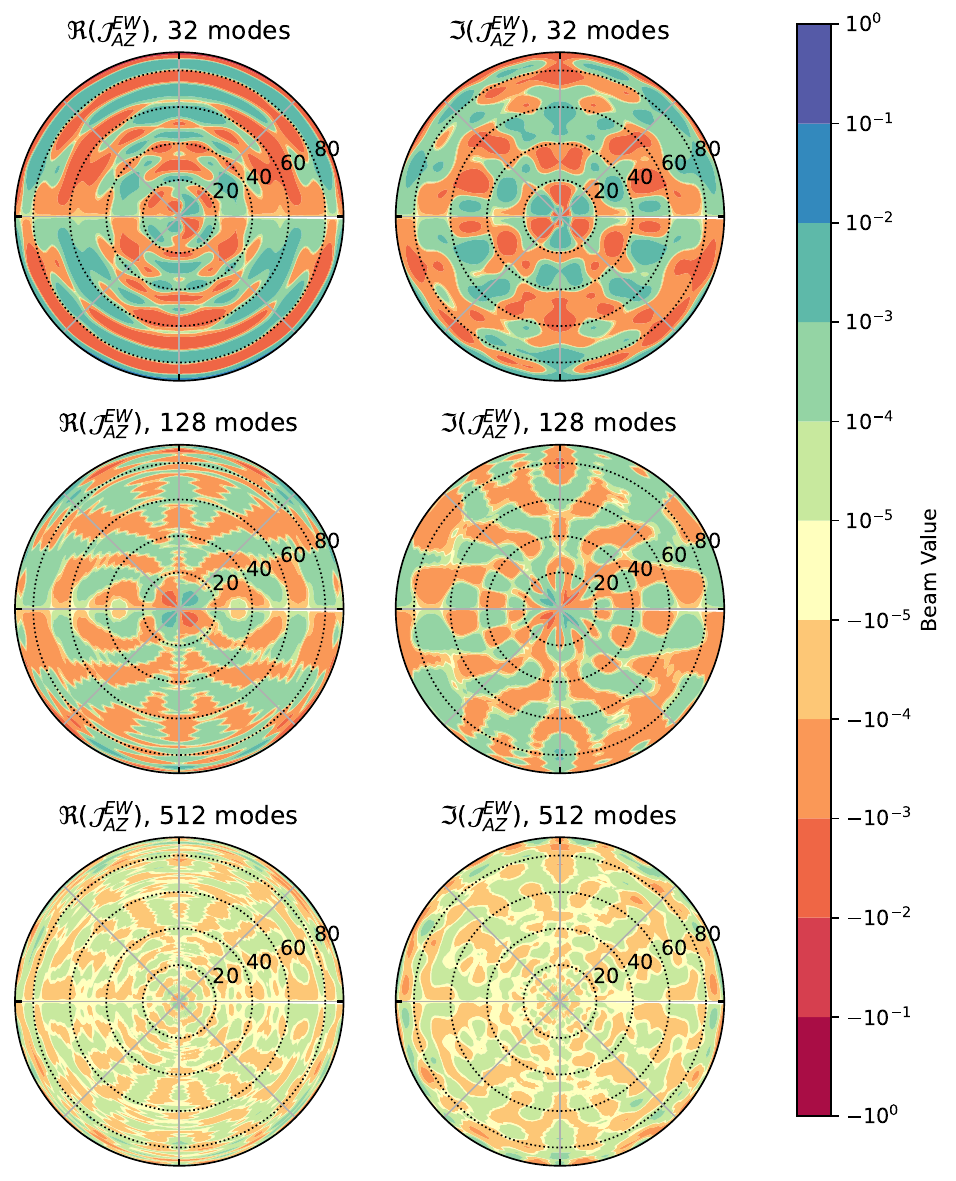}
    \caption{Residuals for the Phase I dipole's response to azimuthally polarized radiation with 32, 128, and 512 basis functions at 150 MHz from top to bottom. The left column shows the real component, and the right shows the imaginary component. With just 32 basis functions, errors are below 1\% of the peak value almost everywhere. }
    \label{fig:dip_sparse}
\end{figure*}

\begin{figure*}
    \includegraphics[width=\linewidth]{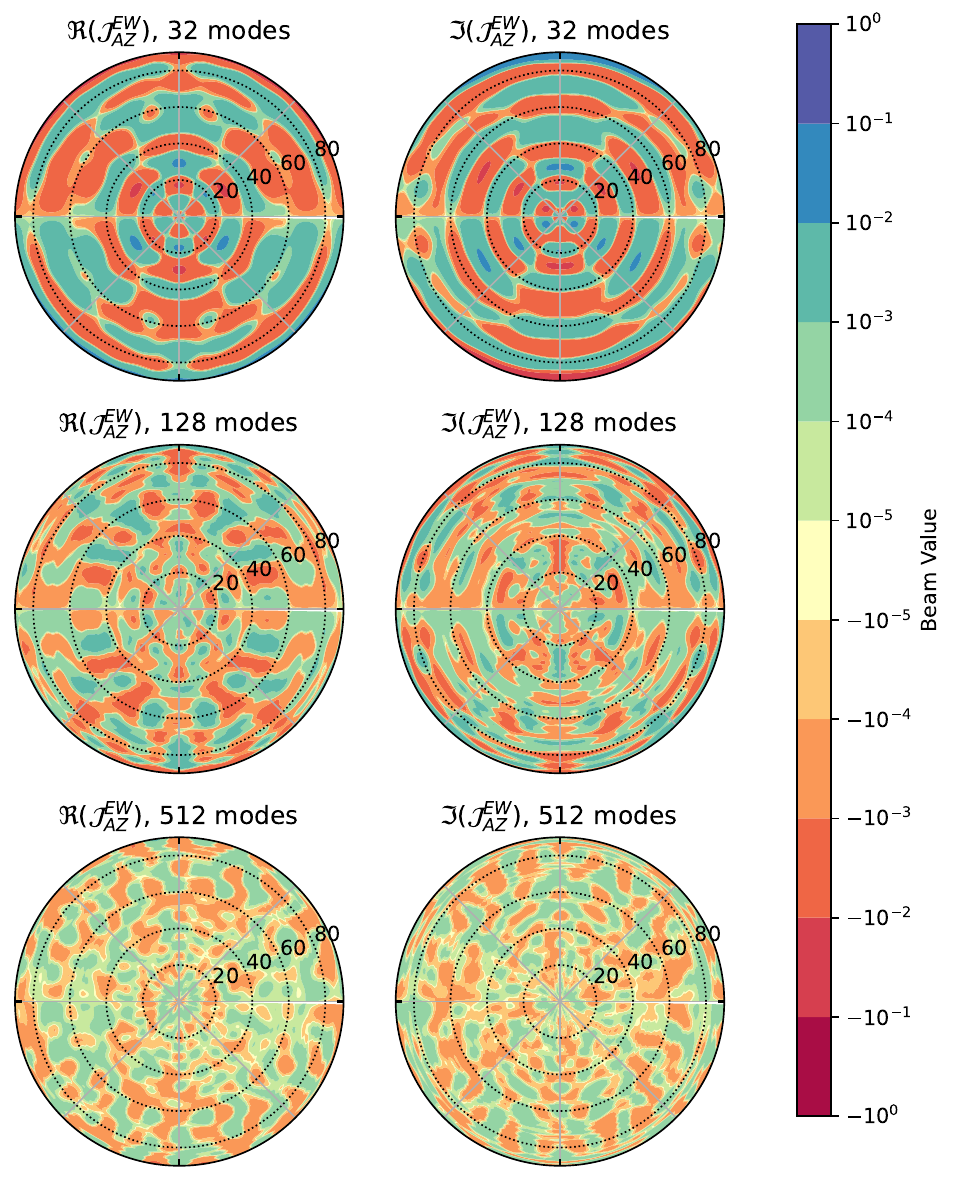}
    
    \caption{Same as Figure \ref{fig:dip_sparse} but for the Vivaldi feed. Performance is similar (though it looks worse due to the discrete binning; cf. Figure \ref{fig:ang_rms}).}
    \label{fig:viv_sparse}
\end{figure*}

\begin{figure*}
    \centering
    \includegraphics[width=\linewidth]{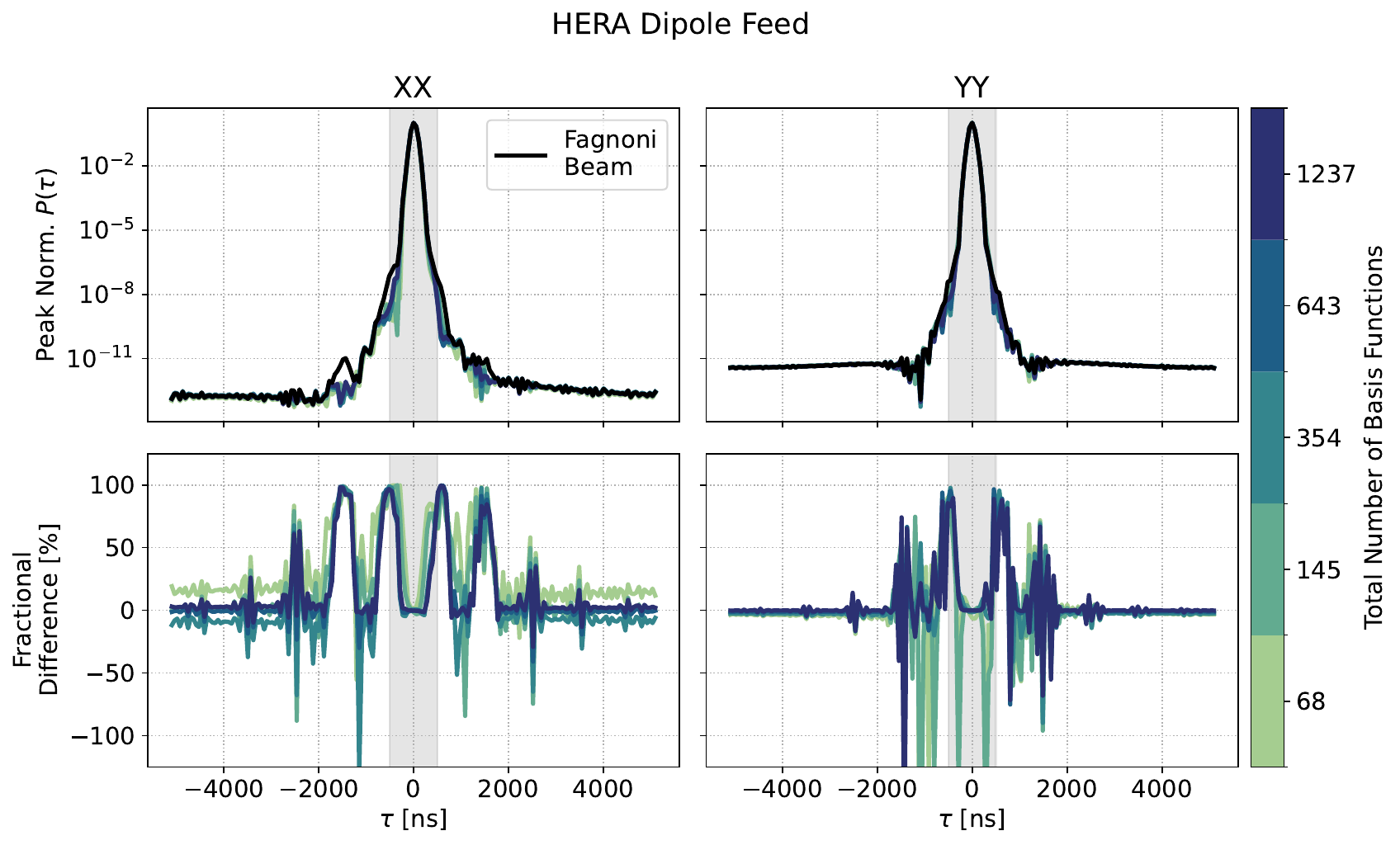}
    \caption{({\it Top}) Delay power spectra ($y$-axis) for the simulations described in section \ref{sec:spectral} using the HERA Dipole feed as a function of delay ($\tau$, $x$-axis) for the XX (left) and YY (right) polarizations.  The solid black lines show the delay power spectra of the visibilities simulated with the CST beam from \citet{Fagnoni2021a}.  The colored lines show the delay power spectra for the simulations using our fits to the CST beam using various numbers of basis functions, $N_\text{coeff}$, as indicated in the colorbar.  All power spectra have been normalized relative to the value of the black line at $\tau = 0$.  The grey shaded region marks the delays less than or equal to the Nyquist delay (500 ns) corresponding to the spectral resolution of the raw Fagnoni beam simulations (1 MHz).  ({\it Bottom}) Fractional difference between the Fagnoni beam and FB simulations for each value of $N_{\text{coeff}}$.  While the delay power spectra shown here correspond to a single 14.6 m EW baseline, the results are representative of all simulated baseline lengths and orientations.}
    % The simulation using the CST is larger in amplitude than the FB fit sims at $|\tau|\lesssim1800$ ns.  This is likely due to interpolation artifacts arising from interpolating the raw Fagnoni beams during visibility simulation (see section \ref{sec:spectral} for details).
    % The delay power spectra for the FB sims are approximately fixed for $N_{\text{coeff}}\geq64$.
    \label{fig:dip-dps-comparison}
\end{figure*}

To demonstrate that our intuition about mode significance holds, we perform sparse fits to the beam using only a subset of the entire basis set considered so far. In each fit, we set a certain number of basis functions to use and pick the basis functions in descending FBES order i.e. we select the $M$ highest-energy basis functions for some $M$ and fit only to those. We then calculate the root-mean-square error (RMSE) of the fit over the whole sky, which is the same as computing $\sqrt{L/N_\text{pix}}$ where $L$ is defined in Equation \ref{eq:lstsq} and $N_\text{pix}$ is the number of pixels in the simulated beam. We show the results in Figure \ref{fig:rmse}, where for each point we increase the number of basis functions by a power of 2. We see that the RMSE decreases roughly as a power law as we include more basis functions. The Vivaldi beam appears consistently less compressible than the Phase I dipole in the sense that we observe a higher RMSE for a given number of basis functions. 

Beam errors manifest as flux and phase errors for sources observed at the location of the beam errors. It is difficult to track these flux and phase errors through the complicated analysis pipelines currently in place for power spectrum estimation. The exact error tolerance for 21-cm power spectrum estimation is thus not obviously determined. \citet{Ewall-Wice2017} suggests it should be better than 1\% everywhere on the sky. Better than $\sim 100\%$ errors in the sidelobes would allow this basis to improve on recent drone-based measurements assuming the sparse selection describes the physical perturbations \citep{echo}.

Figure \ref{fig:ang_rms} shows the RMSE as a function of zenith angle for both feed types and polarizations and a range of sparsity. We find that with 32 modes, there is $\sim 1\%$ agreement in the main lobe and $\sim 10\%$ agreement in the sidelobes. With 512 modes we achieve better than 1\% precision at almost all zenith angles. The residuals from some of the sparse fits are shown in Figures \ref{fig:dip_sparse} and \ref{fig:viv_sparse} for the Phase I dipole and Vivaldi feeds, respectively. For a given level of sparsity, errors generally range by one order of magnitude above and below the RMSE value depending on the location of interest. We explore computational complexity for our Gibbs sampling pipeline in Paper II more thoroughly, however we expect having a beam model with 64-128 parameters is computationally feasible, while 512 modes will be tractable with a more optimized implementation. 

We note that translating the observed errors in this fit to the actual errors produced by this approach in practice is not necessarily straightforward. The actual beam will at best be some perturbation of the simulated beam used in this work, not even including the close-packed array effects, which are known to vary even between different simulation packages on the order of a few percent \citep{Bolli2023}. Perhaps more importantly, the approach in Paper II involves inferring the beam based on the interferometric visibilities, whereas this least-squares fit is more analogous to a problem where one infers the beam based on a holographic map. In other words, this work verifies that the spatial structure of the beam can be captured by this basis with relatively few basis functions, but the errors shown do not necessarily represent the types of errors we may observe with a Bayesian point estimate since the information content of the visibilities will be different than this simulation. Furthermore, our approach will return a full Bayesian posterior, not just a point estimate. Folding uncertainties into a forward modeling procedure should hopefully alleviate some of the difficulties that come with point estimation such as calibration errors \citep{Byrne2021}.

The Fourier-Bessel basis has the advantage that it is sparse enough for our purposes, but probably flexible enough to handle perturbations. An alternative basis can be developed based on singular value decomposition (SVD) of the Jones elements at each frequency. In some preliminary experimentation, we find that this can describe the simulations extremely sparsely, and this is also demonstrated for different instruments in other works \citep{Asad2021, cumner2023}. To remove the dependency on a particular pixelization, we can write the SVD basis functions in terms of our Fourier-Bessel basis, thus developing an approximate SVD basis that smoothly interpolates to any desired spatial position. While this produces an extremely sparse representation of a given simulation, the modes uncovered in the SVD are organized in a nontrivial manner e.g. multiple azimuthal modes describing mixtures of dipolar and quadrupolar structures. This makes it less interpretable and therefore disadvantageous from an instrumental characterization perspective compared to the Fourier-Bessel basis. Since sparsity gains from an SVD approach are significant, we aim to explore this option more extensively in the future. 

\begin{figure*}
    \centering
    \includegraphics[width=\linewidth]{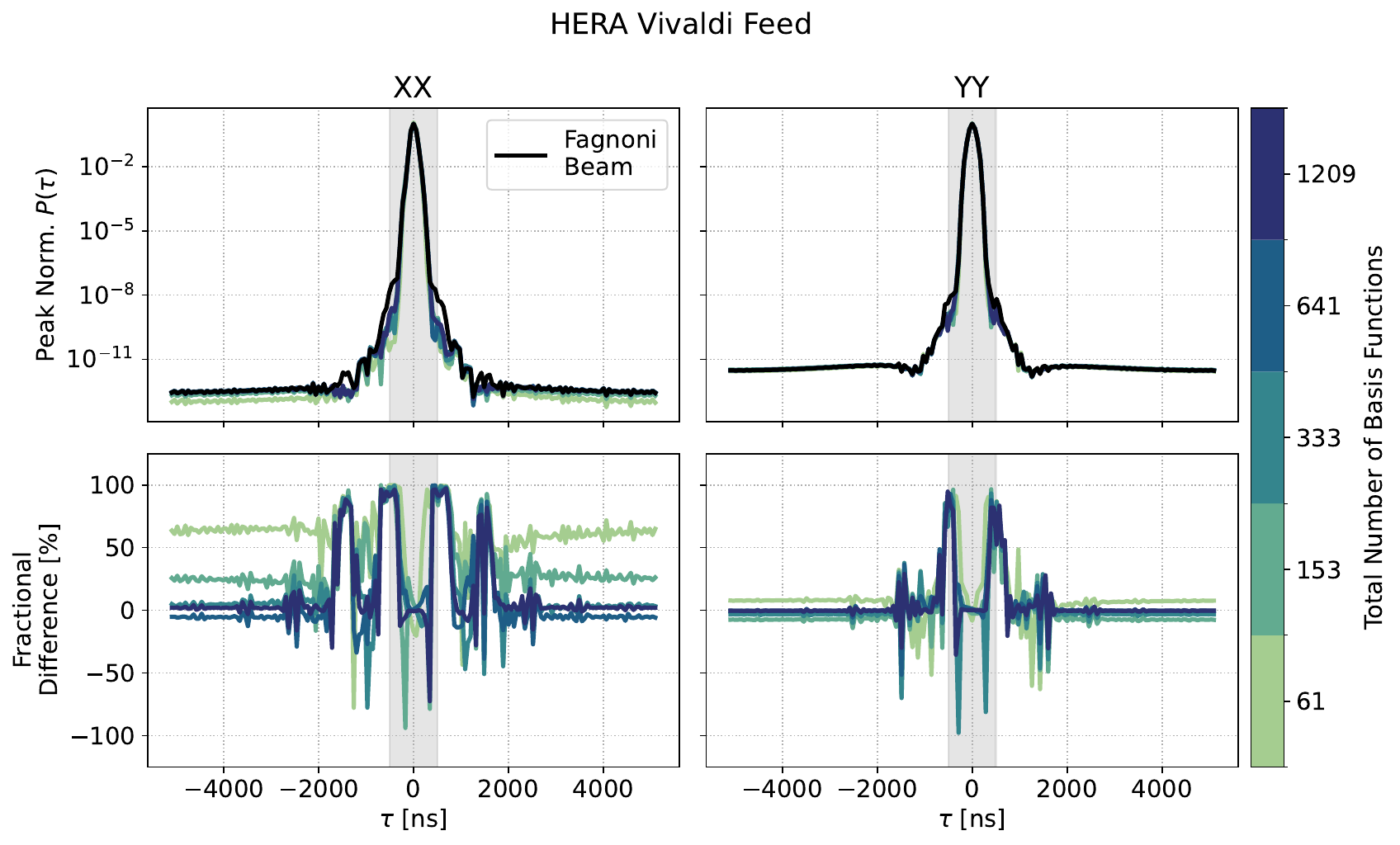}
    \caption{Same as in Figure \ref{fig:dip-dps-comparison} but using the HERA Vivaldi feed from \citet{Fagnoni2021b}.}
    \label{fig:viv-dps-comparison}
\end{figure*}

Spectral structure in the beam errors will greatly magnify their harm, particularly for foreground avoidance strategies that hope to separate the EoR and foreground signals by taking advantage of the spectral smoothness of the foregrounds. In addition, errors in the beam can produce errors in calibration, which can themselves obscure observations of the cosmic reionization signal. Indeed, part of the motivation for a Bayesian approach to beam modeling is to reduce the knock-on effect of beam errors on calibration. We explore the spectral structure of the fit beams in the next section using a suite of simulations, but leave the task of examining effects on calibration to future work. 

\section{Spectral Structure}
\label{sec:spectral}

We investigate any spurious spectral contamination introduced by fitting the beam with the chosen basis.  To quantify this contamination, we compared the delay power spectra \citep{parsons2009, parsons2012} from a suite of visibility simulations which use differing numbers of fit coefficients for the beam.  All other input parameters to the simulations were kept fixed, however, to isolate the effects of changing the beam. 

The delay spectrum is accessed by Fourier transforming (and then squaring) the visibilities along the frequency axis into ``delay" space, so-named because the visibilities in this space represent timelike correlations between the antennas' voltage signals. The intrinsic signal of astrophysical foregrounds is generally confined to delays less than the geometric delay of the baseline, but can be broadened due to instrumental effects (such as a chromatic beam or calibration errors), as well as the choice of spectral tapering function when performing the Fourier transform. Depending on baseline length and other chromatic instrumental effects, delays below a certain scale will be inaccessible without extremely accurate foreground subtraction, while delays greater than this will likely avoid foreground-related contamination \citep{Datta2010, Morales2012, parsons2012, Trott2012}. For HERA, preserving high dynamic range for delays greater than $\sim$200-300 ns is essential to measurements of the 21-cm power spectrum during reionization \citep{parsons2012, HC2023}. Therefore, the primary purpose of this test is to check whether there is a significant decrease in dynamic range above those delays. 

The simulations were performed using \texttt{pyuvsim}\footnote{\url{https://github.com/RadioAstronomySoftwareGroup/pyuvsim}} \citep{pyuvsim}, a high-precision visibility simulator which has been rigorously tested \citep{Lanman2022, Aguirre2022}.  We simulated a selection of baselines from a close-packed hexagonal array layout with various orientations and lengths ($0\leq b\leq 90$ m).  For the sky, we used the 2016 Global Sky Model (GSM, \cite{GSM}), a map of galactic diffuse emission.  For each antenna feed type (dipole or vivaldi), we ran simulations using varied numbers of beam fit coefficients, from 32 to 512 (``FB simulations'').  For a given number of fit coefficients ($N_{\text{coeff}}$), we use the FBES to select the $N_{\text{coeff}}$ modes with the largest FBES amplitudes. We do this independently at each frequency, and take the union of all basis functions selected in this way across the simulated band. Since the top $N_\text{coeff}$ contributing basis functions varies slightly as a function of frequency, this means each selected $N_\text{coeff}$ is less than the number of modes actually used in each instance of the test. Note that this is different in previous sections, where we were only examining one frequency. As a reference against which we can compare the FB simulations, we also ran a simulation for each antenna feed type using the corresponding raw electromagnetic simulations from \citet{Fagnoni2021a, Fagnoni2021b} (``Fagnoni simulations'').  %

% In both Figures \ref{fig:dip-dps-comparison} and \ref{fig:viv-dps-comparison}, we see an excess of power in the Fagnoni simulations (black line) relative to the FB simulations (colored lines) at $|\tau| \lesssim 1800$ ns.  This excess power is likely an interpolation artifact from frequency interpolation of the Fagnoni beams during visibility simulation inside \texttt{pyuvsim}.

This test is greatly complicated by the fact that the reference beam simulations are only available at 1 MHz resolution, which cannot access most of the delays relevant to EoR science. To accommodate a more straight-forward comparison, we fit the Fourier-Bessel beam at 1 MHz resolution and then interpolated the coefficients (as well as the reference beam) during simulation using a cubic spline to a resolution of $\sim97$ kHz to match the HERA spectral resolution. The Nyquist frequency for a 1 MHz bandwidth is 500 ns.  Delays in the range $|\tau|>500$ ns are therefore susceptible to interpolation artifacts regardless of which beam is used. In other words, there is no obvious ground truth of what HERA should observe with this sky model above the Nyquist delay. With this caveat in mind, we list three important observations of the delay spectra from these visibility simulations, and then discuss them in more detail below. The delay spectra are shown in Figures \ref{fig:dip-dps-comparison} and \ref{fig:viv-dps-comparison} for the dipole and Vivaldi feeds, respectively. 

\begin{enumerate}
    \item The Fourier-Bessel beams generally produce slightly smoother visibilities  than the simulated beams of the corresponding feed i.e. there is a loss of power at intermediate delays and beyond, meaning there is no hit to dynamic range.

    \item There is a significant discrepancy below the Nyquist delay, visible in the XX delay spectra, which appear to have pronounced shoulders when using the reference beam. We argue that these are largely an artefact that stems from a complex interplay between spatial interpolation of the beam and the chosen sky model. If this is the case, this structure is spurious and it is desirable that the FB fits should \textit{not} reproduce it.

    \item There is also a significant discrepancy at $|\tau| \sim 1500$ ns in the XX beams, which is above the Nyquist delay. We suspect these are the pronounced shoulders from the reference beams being aliased to a harmonic of the Nyquist delay as a result of the frequency interpolation. This therefore represents high-delay contamination that is happily lacking when using the Fourier-Bessel beams.
\end{enumerate}

%It is therefore plausible that any excess power we see beyond 500 ns in the delay power spectrum of the Fagnoni simulations relative to the Fourier-Bessel simulations is a result of the frequency interpolation of the Fagnoni beams.  We can see potential evidence of this in Figures \ref{fig:dip-dps-comparison} and \ref{fig:viv-dps-comparison}.  For both feed types, there are notable spikes in the fractional difference for all values of $N_{\rm{coeff}}$  at 500 ns and 1500 ns (a harmonic of the 500 ns Nyquist frequency).

%It is important to note that both the raw Fagnoni beams and our FB fit coefficients were required to be interpolated in frequency for our visibility simulations. 
%Interpolating a set of coefficients for a set of analytic functions is in principle quite different from interpolating each pixel individually in the Fagnoni beam simulations, as is done inside \texttt{pyuvsim}.  For the results shown in Figures \ref{fig:dip-dps-comparison} and \ref{fig:viv-dps-comparison}, we used cubic frequency interpolation of the FB fit coefficients.  We similarly used cubic interpolation for the raw Fagnoni beams when running \texttt{pyuvsim}.  
To better understand the choice of interpolation spline, we also ran a \texttt{pyuvsim} simulation using linear frequency interpolation of the Fagnoni beam.  For linear interpolation, interpolation artifacts were highly visible as a set of peaks (duplicates of the peak at $|\tau|\leq 500$ ns) spaced uniformly every 500 ns (the Nyquist frequency for 1 MHz spectral resolution).  While the amplitude of these duplicated peaks decreased with increasing $|\tau|$, the dynamic range between $\tau = 0$ ns and the highest delays was only $\sim10^8$.  Whereas for the cubically interpolated Fagnoni simulations, we see a dynamic range of $\sim10^{11}$ in Figures \ref{fig:dip-dps-comparison} and \ref{fig:viv-dps-comparison} and little evidence of such duplicated low-delay structure.

There are, however, also differences between the Fagnoni beam and FB beam simulations inside the Nyquist delay.  The raw Fagnoni beams were simulated on a rectilinear grid in altitude and azimuth with a 1 deg resolution.  These beams thus also require spatial interpolation during visibility simulation to get the beam value at the location of each source (pixel) in the sky catalog.  Our FB beam fits are analytic and thus interpolate smoothly to any pixelation scheme.  To determine the importance of this spatial interpolation on our results, we also performed a set of simulations using a mock, gridded point source catalog with one point source at the center of each Fagnoni beam pixel at a single LST. 
 In this case, the Fagnoni beams require no spatial interpolation. For this gridded point source catalog, we see almost no discrepancy inside the Nyquist delay.  This suggests that the discrepancies we see for $|\tau|\sim 500$ ns are related to the spatial interpolation of the Fagnoni beam during visibility simulation i.e. it does not represent intrinsic spectral structure of the beam that is subsequently imprinted on any sky model for which it is used. However, since the gridded sky model does not have the same intrinsic spectral structure as the GSM, this additional test by itself leaves open the question as to whether these shoulders are intrinsic beam structure that somehow goes missing when we use the FB beams. By running the simulations of the GSM with an achromatic beam equal to the FB beam at the lowest interpolated frequency in the band, we find that these shoulders are not present, suggesting that the shoulders are not intrinsic to the GSM. In summary, we suspect the large discrepancies between the Fagnoni beam simulations and the best-fit FB beam (see Figures \ref{fig:dip-dps-comparison} and \ref{fig:viv-dps-comparison}) are caused by interpolation issues with the simulated beams that the FB basis successfully avoids, rather than being due to a serious reconstruction error on the part of the FB basis.

Figures \ref{fig:dip-dps-comparison} and \ref{fig:viv-dps-comparison} also show that there is some variation between the FB simulations as a function of $N_{\text{coeff}}$.  Most notably, the dynamic range changes slightly as $N_{\text{coeff}}$ increases.  This is most obvious in the XX polarization results for both feed types.  This effect, however, is quite small and only appears significant in the fractional difference plots because of the low amplitude of the delay power spectrum of the Fagnoni beam simulations. Overall, the dynamic range is large ($\sim10^{13}$) and roughly consistent for both feed types, both polarizations, and all $N_{\text{coeff}}$ values.  This suggests there is minimal spectral structure imposed on the visibilities by the FB beam fits as a function of $N_{\text{coeff}}$.

%{\color{red} JB: I'm not really happy with this section yet.  For $|\tau|>500$ ns we have a compelling argument as to why the sims disagree (Fagnoni beam vs FB fits).  But for $|\tau|<500$ ns, there's also disagreement in the sims.  We see similar disagreement too in the tests I ran with the point sources on/off CST grid centers.  I don't know what we can say about the differences in the region $|\tau|<500$ ns.}

\section{Summary and Conclusion}
\label{sec:conclusion}

We have investigated parametric models of the HERA beams as a means of exploring the addition of primary beam inference in Bayesian 21-cm intensity mapping pipelines. Though we have not specifically investigated it in this work, we expect that allowing for variations in the beam model will help mitigate systematic inference errors in other nuisance parameters such as the direction-independent gains, which is a common problem in pipelines that assume an exact, but incorrect, beam model. Using an analytic basis also simultaneously solves a spatial interpolation problem that arises when using pixel-based beams. Namely, since we choose a spatially smooth basis, the interpolated beam values are also spatially smooth. Since spatial modes couple into spectral modes in 21-cm power spectrum estimation, this enhances the quality of such measurements. 

We found that using a modified Fourier-Bessel basis in an area-preserving projection of the Azimuth-Zenith-Angle coordinate system to the unit disc provided beam models that were relatively compressible for a given desired performance at zenith. Good performance at zenith is a particularly important feature for a drift-scan telescope such as HERA, while compressibility is important to reduce the computational cost of the inference pipeline. To compress the basis, we formed a Fourier-Bessel Energy Spectrum and determined a fixed number of important modes by choosing the strongest contributors to the spectrum. To assess the effectiveness of this compression, we took the least squares fit in the compressed basis and compared fit residuals and root-mean-square error over the sky. Without a full test of how such errors affect e.g. direction-independent calibration, it is difficult to know what constitutes a sufficiently complex beam. However, a baseline test of the basis' effectiveness is to simulate delay spectra using the fit beams with varying number of basis functions, assuming perfect direction-independent calibration, and compare to the same simulation with the simulated beam. 

When we perform this simulation, we find that there is no sign of deleterious spurious spectral structure. For both feeds, performance at high delays greater than 2000 ns is nearly identical to the simulated beams. Performance at intermediate to high delays is within the dynamic range requirements set by the reionization signal. We observed the largest differences for both feed types at intermediate delays surrounding the Nyquist delay of the CST simulation (500 ns) and around 1500 ns. The simulated beams exhibit stronger power at these delays. We suppose this is likely to be an interpolation artefact, since structure at delays beyond the Nyquist delay should not be recovered by a spline interpolation except by coincidence, and various simulations suggest the structure is not an intrinsic structure of the simulated beams or sky model.

Overall, while this study is somewhat limited in scope, we find the results encouraging. Since HERA and other instruments seeking the EoR signal are interferometers, an important extension of this framework will be to examine the compressibility or even general effectiveness of this basis in the presence of mutual coupling between receiving elements. However, results from this work should be directly applicable to global 21-cm signal experiments. Another important effect to study in future work is whether the basis is equally effective at characterizing physical perturbations to the receiving elements due to imperfect design, effects of weather conditions such as high wind, improper soil or ground plane modeling, etc. Despite that these types of errors tend to produce highly nontrivial structure in the beam pattern, the fact that the basic problem is tractable in this framework suggests a way forward for more complicated considerations. Most importantly, the application of this within a Bayesian inference framework will allow the observer to infer what types of perturbations their physical beam exhibits relative to their a priori model. This will then prove to be a powerful tool for enhancing the fidelity of precision measurements of 21-cm reionization signals.

\section*{Acknowledgements}

We are grateful to Aman Chokshi for helpful discussions.

This result is part of a project that has received funding from the European Research Council (ERC) under the European Union's Horizon 2020 research and innovation programme (Grant agreement No. 948764; MJW, JB, PB, KAG). We acknowledge the Science and Technology Facilities Council (STFC) for their support of Eloy de Lera Acedo. We acknowledge use of the following software: 
{\tt matplotlib} \citep{matplotlib}, {\tt numpy} \citep{numpy}, and {\tt scipy} \citep{2020SciPy-NMeth}.

\balance

\section*{Data Availability}

The computer code used to model the beam patterns in this paper is available from \url{https://github.com/HydraRadio/Hydra}.

% BIBLIOGRAPHY
\bibliographystyle{mnras}
\bibliography{beam_param}

\newcommand{\noop}[1]{}
\begin{thebibliography}{}
\makeatletter
\relax
\def\mn@urlcharsother{\let\do\@makeother \do\$\do\&\do\#\do\^\do\_\do\%\do\~}
\def\mn@doi{\begingroup\mn@urlcharsother \@ifnextchar [ {\mn@doi@}
  {\mn@doi@[]}}
\def\mn@doi@[#1]#2{\def\@tempa{#1}\ifx\@tempa\@empty \href
  {http://dx.doi.org/#2} {doi:#2}\else \href {http://dx.doi.org/#2} {#1}\fi
  \endgroup}
\def\mn@eprint#1#2{\mn@eprint@#1:#2::\@nil}
\def\mn@eprint@arXiv#1{\href {http://arxiv.org/abs/#1} {{\tt arXiv:#1}}}
\def\mn@eprint@dblp#1{\href {http://dblp.uni-trier.de/rec/bibtex/#1.xml}
  {dblp:#1}}
\def\mn@eprint@#1:#2:#3:#4\@nil{\def\@tempa {#1}\def\@tempb {#2}\def\@tempc
  {#3}\ifx \@tempc \@empty \let \@tempc \@tempb \let \@tempb \@tempa \fi \ifx
  \@tempb \@empty \def\@tempb {arXiv}\fi \@ifundefined
  {mn@eprint@\@tempb}{\@tempb:\@tempc}{\expandafter \expandafter \csname
  mn@eprint@\@tempb\endcsname \expandafter{\@tempc}}}

\bibitem[\protect\citeauthoryear{{Aguirre} et~al.,}{{Aguirre}
  et~al.}{2022}]{Aguirre2022}
{Aguirre} J.~E.,  et~al., 2022, \mn@doi [\apj] {10.3847/1538-4357/ac32cd},
  \href {https://ui.adsabs.harvard.edu/abs/2022ApJ...924...85A} {924, 85}

\bibitem[\protect\citeauthoryear{Anstey}{Anstey}{2023}]{anstey2023}
Anstey D.,  2023, PhD thesis, University of Cambridge

\bibitem[\protect\citeauthoryear{{Asad} et~al.,}{{Asad}
  et~al.}{2021}]{Asad2021}
{Asad} K.~M.~B.,  et~al., 2021, \mn@doi [\mnras] {10.1093/mnras/stab104}, \href
  {https://ui.adsabs.harvard.edu/abs/2021MNRAS.502.2970A} {502, 2970}

\bibitem[\protect\citeauthoryear{{Barry} \& {Chokshi}}{{Barry} \&
  {Chokshi}}{2022}]{Barry2022}
{Barry} N.,  {Chokshi} A.,  2022, \mn@doi [\apj] {10.3847/1538-4357/ac5903},
  \href {https://ui.adsabs.harvard.edu/abs/2022ApJ...929...64B} {929, 64}

\bibitem[\protect\citeauthoryear{{Barry}, {Hazelton}, {Sullivan}, {Morales}  \&
  {Pober}}{{Barry} et~al.}{2016}]{Barry2016}
{Barry} N.,  {Hazelton} B.,  {Sullivan} I.,  {Morales} M.~F.,   {Pober} J.~C.,
  2016, \mn@doi [\mnras] {10.1093/mnras/stw1380}, \href
  {https://ui.adsabs.harvard.edu/abs/2016MNRAS.461.3135B} {461, 3135}

\bibitem[\protect\citeauthoryear{{Berger} et~al.,}{{Berger}
  et~al.}{2016}]{Berger2016}
{Berger} P.,  et~al., 2016, in {Hall} H.~J.,  {Gilmozzi} R.,   {Marshall}
  H.~K.,  eds,  Society of Photo-Optical Instrumentation Engineers (SPIE)
  Conference Series Vol. 9906, Ground-based and Airborne Telescopes VI. p.
  99060D (\mn@eprint {arXiv} {1607.01473}), \mn@doi{10.1117/12.2233782}

\bibitem[\protect\citeauthoryear{Bolli, Davidson, Labate  \& Wijnholds}{Bolli
  et~al.}{2023}]{Bolli2023}
Bolli P.,  Davidson D.~B.,  Labate M.~G.,   Wijnholds S.~J.,  2023, \mn@doi
  [IEEE Antennas and Wireless Propagation Letters] {10.1109/LAWP.2023.3280169},
  22, 2730

\bibitem[\protect\citeauthoryear{{Bui-Van}, {Craeye}  \& {de Lera
  Acedo}}{{Bui-Van} et~al.}{2017}]{bui2017}
{Bui-Van} H.,  {Craeye} C.,   {de Lera Acedo} E.,  2017, \mn@doi [Experimental
  Astronomy] {10.1007/s10686-017-9565-y}, \href
  {https://ui.adsabs.harvard.edu/abs/2017ExA....44..239B} {44, 239}

\bibitem[\protect\citeauthoryear{{Byrne}}{{Byrne}}{2023}]{Byrne2023}
{Byrne} R.,  2023, \mn@doi [\apj] {10.3847/1538-4357/acac95}, \href
  {https://ui.adsabs.harvard.edu/abs/2023ApJ...943..117B} {943, 117}

\bibitem[\protect\citeauthoryear{{Byrne} et~al.,}{{Byrne}
  et~al.}{2019}]{Byrne2019}
{Byrne} R.,  et~al., 2019, \mn@doi [\apj] {10.3847/1538-4357/ab107d}, \href
  {https://ui.adsabs.harvard.edu/abs/2019ApJ...875...70B} {875, 70}

\bibitem[\protect\citeauthoryear{{Byrne}, {Morales}, {Hazelton}  \&
  {Wilensky}}{{Byrne} et~al.}{2021}]{Byrne2021}
{Byrne} R.,  {Morales} M.~F.,  {Hazelton} B.~J.,   {Wilensky} M.,  2021,
  \mn@doi [\mnras] {10.1093/mnras/stab647}, \href
  {https://ui.adsabs.harvard.edu/abs/2021MNRAS.503.2457B} {503, 2457}

\bibitem[\protect\citeauthoryear{{Byrne}, {Morales}, {Hazelton}, {Sullivan}  \&
  {Barry}}{{Byrne} et~al.}{2022}]{Byrne2022}
{Byrne} R.~L.,  {Morales} M.~F.,  {Hazelton} B.,  {Sullivan} I.,   {Barry} N.,
  2022, \mn@doi [\pasa] {10.1017/pasa.2022.21}, \href
  {https://ui.adsabs.harvard.edu/abs/2022PASA...39...23B} {39, e023}

\bibitem[\protect\citeauthoryear{{CHIME Collaboration} et~al.,}{{CHIME
  Collaboration} et~al.}{2022}]{CHIME2022}
{CHIME Collaboration} et~al., 2022, \mn@doi [\apjs] {10.3847/1538-4365/ac6fd9},
  \href {https://ui.adsabs.harvard.edu/abs/2022ApJS..261...29C} {261, 29}

\bibitem[\protect\citeauthoryear{{Charles}, {Kern}, {Bernardi}, {Bester},
  {Smirnov}, {Fagnoni}  \& {Acedo}}{{Charles} et~al.}{2023}]{Charles2023}
{Charles} N.,  {Kern} N.,  {Bernardi} G.,  {Bester} L.,  {Smirnov} O.,
  {Fagnoni} N.,   {Acedo} E. d.~L.,  2023, \mn@doi [\mnras]
  {10.1093/mnras/stad1046}, \href
  {https://ui.adsabs.harvard.edu/abs/2023MNRAS.522.1009C} {522, 1009}

\bibitem[\protect\citeauthoryear{{Choudhuri}, {Bull}  \& {Garsden}}{{Choudhuri}
  et~al.}{2021}]{Choudhuri+21}
{Choudhuri} S.,  {Bull} P.,   {Garsden} H.,  2021, \mn@doi [\mnras]
  {10.1093/mnras/stab1795}, \href
  {https://ui.adsabs.harvard.edu/abs/2021MNRAS.506.2066C} {506, 2066}

\bibitem[\protect\citeauthoryear{{Clemens} \& {Weiland}}{{Clemens} \&
  {Weiland}}{2001}]{Clemens2001}
{Clemens} M.,  {Weiland} T.,  2001, \mn@doi [] {10.2528/PIER00080103}, 32, 65

\bibitem[\protect\citeauthoryear{{Cox}, {Parsons}, {Dillon}, {Ewall-Wice}  \&
  {Pascua}}{{Cox} et~al.}{2023}]{Cox2023}
{Cox} T.~A.,  {Parsons} A.~R.,  {Dillon} J.~S.,  {Ewall-Wice} A.,   {Pascua}
  R.,  2023, \mn@doi [arXiv e-prints] {10.48550/arXiv.2311.01422}, \href
  {https://ui.adsabs.harvard.edu/abs/2023arXiv231101422C} {p. arXiv:2311.01422}

\bibitem[\protect\citeauthoryear{{Cumner}, {Pieterse}, {De Villiers}  \& {de
  Lera Acedo}}{{Cumner} et~al.}{2023}]{cumner2023}
{Cumner} J.,  {Pieterse} C.,  {De Villiers} D.,   {de Lera Acedo} E.,  2023,
  \mn@doi [arXiv e-prints] {10.48550/arXiv.2311.07392}, \href
  {https://ui.adsabs.harvard.edu/abs/2023arXiv231107392C} {p. arXiv:2311.07392}

\bibitem[\protect\citeauthoryear{{Datta}, {Bowman}  \& {Carilli}}{{Datta}
  et~al.}{2010}]{Datta2010}
{Datta} A.,  {Bowman} J.~D.,   {Carilli} C.~L.,  2010, \mn@doi [\apj]
  {10.1088/0004-637X/724/1/526}, \href
  {https://ui.adsabs.harvard.edu/abs/2010ApJ...724..526D} {724, 526}

\bibitem[\protect\citeauthoryear{{DeBoer} et~al.,}{{DeBoer}
  et~al.}{2017}]{deBoer2017}
{DeBoer} D.~R.,  et~al., 2017, \mn@doi [\pasp]
  {10.1088/1538-3873/129/974/045001}, \href
  {https://ui.adsabs.harvard.edu/abs/2017PASP..129d5001D} {129, 045001}

\bibitem[\protect\citeauthoryear{{Dillon} et~al.,}{{Dillon}
  et~al.}{2020}]{Dillon2020}
{Dillon} J.~S.,  et~al., 2020, \mn@doi [\mnras] {10.1093/mnras/staa3001}, \href
  {https://ui.adsabs.harvard.edu/abs/2020MNRAS.499.5840D} {499, 5840}

\bibitem[\protect\citeauthoryear{{Ewall-Wice}, {Dillon}, {Liu}  \&
  {Hewitt}}{{Ewall-Wice} et~al.}{2017}]{Ewall-Wice2017}
{Ewall-Wice} A.,  {Dillon} J.~S.,  {Liu} A.,   {Hewitt} J.,  2017, \mn@doi
  [\mnras] {10.1093/mnras/stx1221}, \href
  {https://ui.adsabs.harvard.edu/abs/2017MNRAS.470.1849E} {470, 1849}

\bibitem[\protect\citeauthoryear{{Ewall-Wice}, {Dillon}, {Gehlot}, {Parsons},
  {Cox}  \& {Jacobs}}{{Ewall-Wice} et~al.}{2022}]{AEW2022}
{Ewall-Wice} A.,  {Dillon} J.~S.,  {Gehlot} B.,  {Parsons} A.,  {Cox} T.,
  {Jacobs} D.~C.,  2022, \mn@doi [\apj] {10.3847/1538-4357/ac87b3}, \href
  {https://ui.adsabs.harvard.edu/abs/2022ApJ...938..151E} {938, 151}

\bibitem[\protect\citeauthoryear{{Fagnoni}, {de Lera Acedo}, {Drought},
  {DeBoer}, {Riley}, {Razavi-Ghods}, {Carey}  \& {Parsons}}{{Fagnoni}
  et~al.}{2021a}]{Fagnoni2021b}
{Fagnoni} N.,  {de Lera Acedo} E.,  {Drought} N.,  {DeBoer} D.~R.,  {Riley} D.,
   {Razavi-Ghods} N.,  {Carey} S.,   {Parsons} A.~R.,  2021a, \mn@doi [IEEE
  Transactions on Antennas and Propagation] {10.1109/TAP.2021.3083788}, \href
  {https://ui.adsabs.harvard.edu/abs/2021ITAP...69.8143F} {69, 8143}

\bibitem[\protect\citeauthoryear{{Fagnoni} et~al.,}{{Fagnoni}
  et~al.}{2021b}]{Fagnoni2021a}
{Fagnoni} N.,  et~al., 2021b, \mn@doi [\mnras] {10.1093/mnras/staa3268}, \href
  {https://ui.adsabs.harvard.edu/abs/2021MNRAS.500.1232F} {500, 1232}

\bibitem[\protect\citeauthoryear{{Gelman}, {Carlin}, {Stern}, {Dunson},
  {Vehtari}  \& {Rubin}}{{Gelman} et~al.}{2013}]{Gelman2013}
{Gelman} A.,  {Carlin} J.~B.,  {Stern} H.~S.,  {Dunson} D.~B.,  {Vehtari} A.,
  {Rubin} D.~B.,  2013, {Bayesian Data Analysis}

\bibitem[\protect\citeauthoryear{{HERA Collaboration} et~al.,}{{HERA
  Collaboration} et~al.}{2023}]{HC2023}
{HERA Collaboration} et~al., 2023, \mn@doi [\apj] {10.3847/1538-4357/acaf50},
  \href {https://ui.adsabs.harvard.edu/abs/2023ApJ...945..124H} {945, 124}

\bibitem[\protect\citeauthoryear{{Hazelton}, {Jacobs}, {Pober}  \&
  {Beardsley}}{{Hazelton} et~al.}{2017}]{Hazelton2017}
{Hazelton} B.~J.,  {Jacobs} D.~C.,  {Pober} J.~C.,   {Beardsley} A.~P.,  2017,
  \mn@doi [The Journal of Open Source Software] {10.21105/joss.00140}, \href
  {https://ui.adsabs.harvard.edu/abs/2017JOSS....2..140H} {2, 140}

\bibitem[\protect\citeauthoryear{{Hunter}}{{Hunter}}{2007}]{matplotlib}
{Hunter} J.~D.,  2007, Computing in Science Engineering, 9, 90

\bibitem[\protect\citeauthoryear{{Iheanetu}, {Girard}, {Smirnov}, {Asad}, {de
  Villiers}, {Thorat}, {Makhathini}  \& {Perley}}{{Iheanetu}
  et~al.}{2019}]{VLA2019}
{Iheanetu} K.,  {Girard} J.~N.,  {Smirnov} O.,  {Asad} K.~M.~B.,  {de Villiers}
  M.,  {Thorat} K.,  {Makhathini} S.,   {Perley} R.~A.,  2019, \mn@doi [\mnras]
  {10.1093/mnras/stz702}, \href
  {https://ui.adsabs.harvard.edu/abs/2019MNRAS.485.4107I} {485, 4107}

\bibitem[\protect\citeauthoryear{{Jackson}}{{Jackson}}{1998}]{jackson}
{Jackson} J.~D.,  1998, {Classical Electrodynamics, 3rd Edition}.
John Wiley \& Sons, Inc

\bibitem[\protect\citeauthoryear{{Jacobs} et~al.,}{{Jacobs}
  et~al.}{2017}]{echo}
{Jacobs} D.~C.,  et~al., 2017, \mn@doi [\pasp] {10.1088/1538-3873/aa56b9},
  \href {https://ui.adsabs.harvard.edu/abs/2017PASP..129c5002J} {129, 035002}

\bibitem[\protect\citeauthoryear{{Kennedy}, {Bull}, {Wilensky}, {Burba}  \&
  {Choudhuri}}{{Kennedy} et~al.}{2023}]{Kennedy2023}
{Kennedy} F.,  {Bull} P.,  {Wilensky} M.~J.,  {Burba} J.,   {Choudhuri} S.,
  2023, \mn@doi [\apjs] {10.3847/1538-4365/acc324}, \href
  {https://ui.adsabs.harvard.edu/abs/2023ApJS..266...23K} {266, 23}

\bibitem[\protect\citeauthoryear{{Kim} et~al.,}{{Kim} et~al.}{2022}]{Kim2022}
{Kim} H.,  et~al., 2022, \mn@doi [\apj] {10.3847/1538-4357/ac9eaf}, \href
  {https://ui.adsabs.harvard.edu/abs/2022ApJ...941..207K} {941, 207}

\bibitem[\protect\citeauthoryear{{Kim} et~al.,}{{Kim} et~al.}{2023}]{Kim2023}
{Kim} H.,  et~al., 2023, \mn@doi [\apj] {10.3847/1538-4357/ace35e}, \href
  {https://ui.adsabs.harvard.edu/abs/2023ApJ...953..136K} {953, 136}

\bibitem[\protect\citeauthoryear{{Lanman}, {Hazelton}, {Jacobs}, {Kolopanis},
  {Pober}, {Aguirre}  \& {Thyagarajan}}{{Lanman} et~al.}{2019}]{pyuvsim}
{Lanman} A.,  {Hazelton} B.,  {Jacobs} D.,  {Kolopanis} M.,  {Pober} J.,
  {Aguirre} J.,   {Thyagarajan} N.,  2019, \mn@doi [The Journal of Open Source
  Software] {10.21105/joss.01234}, \href
  {https://ui.adsabs.harvard.edu/abs/2019JOSS....4.1234L} {4, 1234}

\bibitem[\protect\citeauthoryear{{Lanman}, {Murray}  \& {Jacobs}}{{Lanman}
  et~al.}{2022}]{Lanman2022}
{Lanman} A.~E.,  {Murray} S.~G.,   {Jacobs} D.~C.,  2022, \mn@doi [\apjs]
  {10.3847/1538-4365/ac45fd}, \href
  {https://ui.adsabs.harvard.edu/abs/2022ApJS..259...22L} {259, 22}

\bibitem[\protect\citeauthoryear{{Line} et~al.,}{{Line}
  et~al.}{2018}]{Line2018}
{Line} J.~L.~B.,  et~al., 2018, \mn@doi [\pasa] {10.1017/pasa.2018.30}, \href
  {https://ui.adsabs.harvard.edu/abs/2018PASA...35...45L} {35, e045}

\bibitem[\protect\citeauthoryear{{Liu} \& {Shaw}}{{Liu} \&
  {Shaw}}{2020}]{liu-shaw}
{Liu} A.,  {Shaw} J.~R.,  2020, \mn@doi [\pasp] {10.1088/1538-3873/ab5bfd},
  \href {https://ui.adsabs.harvard.edu/abs/2020PASP..132f2001L} {132, 062001}

\bibitem[\protect\citeauthoryear{{Lochner}, {Natarajan}, {Zwart}, {Smirnov},
  {Bassett}, {Oozeer}  \& {Kunz}}{{Lochner} et~al.}{2015}]{Lochner2015}
{Lochner} M.,  {Natarajan} I.,  {Zwart} J. T.~L.,  {Smirnov} O.,  {Bassett}
  B.~A.,  {Oozeer} N.,   {Kunz} M.,  2015, \mn@doi [\mnras]
  {10.1093/mnras/stv679}, \href
  {https://ui.adsabs.harvard.edu/abs/2015MNRAS.450.1308L} {450, 1308}

\bibitem[\protect\citeauthoryear{Maaskant, Ivashina, Wijnholds  \&
  Warnick}{Maaskant et~al.}{2012}]{Maaskant2012}
Maaskant R.,  Ivashina M.~V.,  Wijnholds S.~J.,   Warnick K.~F.,  2012, \mn@doi
  [IEEE Transactions on Antennas and Propagation] {10.1109/TAP.2012.2201104},
  60, 3614

\bibitem[\protect\citeauthoryear{Mesinger}{Mesinger}{2019}]{mesinger}
Mesinger A.,  ed. 2019, The Cosmic 21-cm Revolution.
2514-3433, IOP Publishing, \mn@doi{10.1088/2514-3433/ab4a73}, \url
  {https://dx.doi.org/10.1088/2514-3433/ab4a73}

\bibitem[\protect\citeauthoryear{{Morales} \& {Hewitt}}{{Morales} \&
  {Hewitt}}{2004}]{morales-hewitt}
{Morales} M.~F.,  {Hewitt} J.,  2004, \mn@doi [\apj] {10.1086/424437}, \href
  {https://ui.adsabs.harvard.edu/abs/2004ApJ...615....7M} {615, 7}

\bibitem[\protect\citeauthoryear{{Morales}, {Hazelton}, {Sullivan}  \&
  {Beardsley}}{{Morales} et~al.}{2012}]{Morales2012}
{Morales} M.~F.,  {Hazelton} B.,  {Sullivan} I.,   {Beardsley} A.,  2012,
  \mn@doi [\apj] {10.1088/0004-637X/752/2/137}, \href
  {https://ui.adsabs.harvard.edu/abs/2012ApJ...752..137M} {752, 137}

\bibitem[\protect\citeauthoryear{{Morales}, {Beardsley}, {Pober}, {Barry},
  {Hazelton}, {Jacobs}  \& {Sullivan}}{{Morales} et~al.}{2019}]{Morales2019}
{Morales} M.~F.,  {Beardsley} A.,  {Pober} J.,  {Barry} N.,  {Hazelton} B.,
  {Jacobs} D.,   {Sullivan} I.,  2019, \mn@doi [\mnras]
  {10.1093/mnras/sty2844}, \href
  {https://ui.adsabs.harvard.edu/abs/2019MNRAS.483.2207M} {483, 2207}

\bibitem[\protect\citeauthoryear{{Nasirudin}, {Prelogovic}, {Murray},
  {Mesinger}  \& {Bernardi}}{{Nasirudin} et~al.}{2022}]{Nasirudin2022}
{Nasirudin} A.,  {Prelogovic} D.,  {Murray} S.~G.,  {Mesinger} A.,   {Bernardi}
  G.,  2022, \mn@doi [\mnras] {10.1093/mnras/stac1588}, \href
  {https://ui.adsabs.harvard.edu/abs/2022MNRAS.514.4655N} {514, 4655}

\bibitem[\protect\citeauthoryear{{Nunhokee} et~al.,}{{Nunhokee}
  et~al.}{2020}]{2020ApJ...897....5N}
{Nunhokee} C.~D.,  et~al., 2020, \mn@doi [\apj] {10.3847/1538-4357/ab9634},
  \href {https://ui.adsabs.harvard.edu/abs/2020ApJ...897....5N} {897, 5}

\bibitem[\protect\citeauthoryear{Orosz, Dillon, Ewall-Wice, Parsons  \&
  Thyagarajan}{Orosz et~al.}{2019}]{Orosz_2019}
Orosz N.,  Dillon J.~S.,  Ewall-Wice A.,  Parsons A.~R.,   Thyagarajan N.,
  2019, \mn@doi [Monthly Notices of the Royal Astronomical Society]
  {10.1093/mnras/stz1287}, 487, 537

\bibitem[\protect\citeauthoryear{{Paciga} et~al.,}{{Paciga}
  et~al.}{2011}]{Paciga2011}
{Paciga} G.,  et~al., 2011, \mn@doi [\mnras]
  {10.1111/j.1365-2966.2011.18208.x}, \href
  {https://ui.adsabs.harvard.edu/abs/2011MNRAS.413.1174P} {413, 1174}

\bibitem[\protect\citeauthoryear{{Parsons} \& {Backer}}{{Parsons} \&
  {Backer}}{2009}]{parsons2009}
{Parsons} A.~R.,  {Backer} D.~C.,  2009, \mn@doi [\aj]
  {10.1088/0004-6256/138/1/219}, \href
  {https://ui.adsabs.harvard.edu/abs/2009AJ....138..219P} {138, 219}

\bibitem[\protect\citeauthoryear{{Parsons}, {Pober}, {Aguirre}, {Carilli},
  {Jacobs}  \& {Moore}}{{Parsons} et~al.}{2012}]{parsons2012}
{Parsons} A.~R.,  {Pober} J.~C.,  {Aguirre} J.~E.,  {Carilli} C.~L.,  {Jacobs}
  D.~C.,   {Moore} D.~F.,  2012, \mn@doi [\apj] {10.1088/0004-637X/756/2/165},
  \href {https://ui.adsabs.harvard.edu/abs/2012ApJ...756..165P} {756, 165}

\bibitem[\protect\citeauthoryear{{Pober} et~al.,}{{Pober}
  et~al.}{2012}]{2012AJ....143...53P}
{Pober} J.~C.,  et~al., 2012, \mn@doi [\aj] {10.1088/0004-6256/143/2/53}, \href
  {https://ui.adsabs.harvard.edu/abs/2012AJ....143...53P} {143, 53}

\bibitem[\protect\citeauthoryear{{Pupillo} et~al.,}{{Pupillo}
  et~al.}{2015}]{2015ExA....39..405P}
{Pupillo} G.,  et~al., 2015, \mn@doi [Experimental Astronomy]
  {10.1007/s10686-015-9456-z}, \href
  {https://ui.adsabs.harvard.edu/abs/2015ExA....39..405P} {39, 405}

\bibitem[\protect\citeauthoryear{Santos et~al.,}{Santos
  et~al.}{2017}]{santos2017meerklass}
Santos M.~G.,  et~al., 2017, MeerKLASS: MeerKAT Large Area Synoptic Survey
  (\mn@eprint {arXiv} {1709.06099})

\bibitem[\protect\citeauthoryear{{Sekhar}, {Jagannathan}, {Kirk}, {Bhatnagar}
  \& {Taylor}}{{Sekhar} et~al.}{2022}]{Sekhar2022}
{Sekhar} S.,  {Jagannathan} P.,  {Kirk} B.,  {Bhatnagar} S.,   {Taylor} R.,
  2022, \mn@doi [\aj] {10.3847/1538-3881/ac41c4}, \href
  {https://ui.adsabs.harvard.edu/abs/2022AJ....163...87S} {163, 87}

\bibitem[\protect\citeauthoryear{{Shaw}, {Sigurdson}, {Sitwell}, {Stebbins}  \&
  {Pen}}{{Shaw} et~al.}{2015}]{Shaw2015}
{Shaw} J.~R.,  {Sigurdson} K.,  {Sitwell} M.,  {Stebbins} A.,   {Pen} U.-L.,
  2015, \mn@doi [\prd] {10.1103/PhysRevD.91.083514}, \href
  {https://ui.adsabs.harvard.edu/abs/2015PhRvD..91h3514S} {91, 083514}

\bibitem[\protect\citeauthoryear{{Sims}, {Pober}  \& {Sievers}}{{Sims}
  et~al.}{2022a}]{BayesCal1}
{Sims} P.~H.,  {Pober} J.~C.,   {Sievers} J.~L.,  2022a, \mn@doi [\mnras]
  {10.1093/mnras/stac1861}, \href
  {https://ui.adsabs.harvard.edu/abs/2022MNRAS.517..910S} {517, 910}

\bibitem[\protect\citeauthoryear{{Sims}, {Pober}  \& {Sievers}}{{Sims}
  et~al.}{2022b}]{BayesCal2}
{Sims} P.~H.,  {Pober} J.~C.,   {Sievers} J.~L.,  2022b, \mn@doi [\mnras]
  {10.1093/mnras/stac1749}, \href
  {https://ui.adsabs.harvard.edu/abs/2022MNRAS.517..935S} {517, 935}

\bibitem[\protect\citeauthoryear{{Sims} et~al.,}{{Sims}
  et~al.}{2023}]{Sims2023}
{Sims} P.~H.,  et~al., 2023, \mn@doi [\mnras] {10.1093/mnras/stad610}, \href
  {https://ui.adsabs.harvard.edu/abs/2023MNRAS.521.3273S} {521, 3273}

\bibitem[\protect\citeauthoryear{{Thyagarajan}, {Parsons}, {DeBoer}, {Bowman},
  {Ewall-Wice}, {Neben}  \& {Patra}}{{Thyagarajan}
  et~al.}{2016}]{Thyagarajan2016}
{Thyagarajan} N.,  {Parsons} A.~R.,  {DeBoer} D.~R.,  {Bowman} J.~D.,
  {Ewall-Wice} A.~M.,  {Neben} A.~R.,   {Patra} N.,  2016, \mn@doi [\apj]
  {10.3847/0004-637X/825/1/9}, \href
  {https://ui.adsabs.harvard.edu/abs/2016ApJ...825....9T} {825, 9}

\bibitem[\protect\citeauthoryear{{Tingay} et~al.,}{{Tingay}
  et~al.}{2013}]{Tingay2013}
{Tingay} S.~J.,  et~al., 2013, \mn@doi [\pasa] {10.1017/pasa.2012.007}, \href
  {https://ui.adsabs.harvard.edu/abs/2013PASA...30....7T} {30, e007}

\bibitem[\protect\citeauthoryear{{Trott}, {Wayth}  \& {Tingay}}{{Trott}
  et~al.}{2012}]{Trott2012}
{Trott} C.~M.,  {Wayth} R.~B.,   {Tingay} S.~J.,  2012, \mn@doi [\apj]
  {10.1088/0004-637X/757/1/101}, \href
  {https://ui.adsabs.harvard.edu/abs/2012ApJ...757..101T} {757, 101}

\bibitem[\protect\citeauthoryear{{Virone} et~al.,}{{Virone}
  et~al.}{2014}]{2014IAWPL..13..169V}
{Virone} G.,  et~al., 2014, \mn@doi [IEEE Antennas and Wireless Propagation
  Letters] {10.1109/LAWP.2014.2298250}, \href
  {https://ui.adsabs.harvard.edu/abs/2014IAWPL..13..169V} {13, 169}

\bibitem[\protect\citeauthoryear{{Virtanen} et~al.,}{{Virtanen}
  et~al.}{2020}]{2020SciPy-NMeth}
{Virtanen} P.,  et~al., 2020, \mn@doi [Nature Methods]
  {https://doi.org/10.1038/s41592-019-0686-2}, \href {https://rdcu.be/b08Wh}
  {17, 261}

\bibitem[\protect\citeauthoryear{{Weiland}}{{Weiland}}{1977}]{Weiland1977}
{Weiland} T.,  1977, Archiv Elektronik und Uebertragungstechnik, \href
  {https://ui.adsabs.harvard.edu/abs/1977ArElU..31..116W} {31, 116}

\bibitem[\protect\citeauthoryear{Yatawatta}{Yatawatta}{2018}]{Yatawatta2018}
Yatawatta S.,  2018, in 2018 IEEE International Conference on Acoustics, Speech
  and Signal Processing (ICASSP). pp 3489--3493,
  \mn@doi{10.1109/ICASSP.2018.8462230}

\bibitem[\protect\citeauthoryear{Young, Maaskant, Ivashina, de Villiers  \&
  Davidson}{Young et~al.}{2013}]{Young2013}
Young A.,  Maaskant R.,  Ivashina M.~V.,  de Villiers D. I.~L.,   Davidson
  D.~B.,  2013, \mn@doi [IEEE Transactions on Antennas and Propagation]
  {10.1109/TAP.2013.2239954}, 61, 2466

\bibitem[\protect\citeauthoryear{{Zheng} et~al.,}{{Zheng} et~al.}{2017}]{GSM}
{Zheng} H.,  et~al., 2017, \mn@doi [\mnras] {10.1093/mnras/stw2525}, \href
  {https://ui.adsabs.harvard.edu/abs/2017MNRAS.464.3486Z} {464, 3486}

\bibitem[\protect\citeauthoryear{de Lera~Acedo, Craeye, Razavi-Ghods  \&
  González-Ovejero}{de~Lera~Acedo et~al.}{2013}]{deLeraAcedo13}
de Lera~Acedo E.,  Craeye C.,  Razavi-Ghods N.,   González-Ovejero D.,  2013,
  in 2013 International Conference on Electromagnetics in Advanced Applications
  (ICEAA). pp 1182--1185, \mn@doi{10.1109/ICEAA.2013.6632431}

\bibitem[\protect\citeauthoryear{{van Haarlem} et~al.,}{{van Haarlem}
  et~al.}{2013}]{LOFAR}
{van Haarlem} M.~P.,  et~al., 2013, \mn@doi [\aap]
  {10.1051/0004-6361/201220873}, \href
  {https://ui.adsabs.harvard.edu/abs/2013A&A...556A...2V} {556, A2}

\bibitem[\protect\citeauthoryear{{van der Walt}, {Colbert}  \&
  {Varoquaux}}{{van der Walt} et~al.}{2011}]{numpy}
{van der Walt} S.,  {Colbert} S.~C.,   {Varoquaux} G.,  2011, Computing in
  Science Engineering, 13, 22

\makeatother
\end{thebibliography}
\label{lastpage}
\end{document}